\documentclass[a4paper,12pt]{article}

\pdfoutput=1
\usepackage{amsmath,amssymb,booktabs}
\usepackage{epsf,epsfig}
\usepackage{psfrag}
\usepackage{cite}
\usepackage{graphics,subfigure}
\usepackage{mathrsfs}
\usepackage{bbm}
\usepackage{slashed}

\usepackage{multirow}
\usepackage{listings}

\allowdisplaybreaks

\def\no{\nonumber}

\def\MSbar{\ensuremath{\overline{\text{MS}}}}

\def\gev{{\rm GeV}}
\def\mev{{\rm MeV}}

\def\be{\begin{equation}}
\def\ee{\end{equation}}
\def\beq{\begin{eqnarray}}
\def\eeq{\end{eqnarray}}

\def\eps{\epsilon}

\def\DB0{\partial B_0}

\def\Cl2{\mbox{Cl}_2}

\newcommand{\al}{\alpha}
\newcommand{\bet}{\beta}
\newcommand{\gam}{\gamma}
\newcommand{\sig}{\sigma}
\newcommand{\del}{\delta}
\newcommand{\nn}{\nonumber}
\newcommand{\alsfpi}{\frac{{\al}_s}{{4\pi}}}
\newcommand{\alsfpihat}{\frac{{\hat\al}_s}{{4\pi}}}
\newcommand{\alshat}{\hat{\al}_s}
\newcommand{\Lamqcd}{\Lambda_\text{QCD}}
\newcommand{\msbar}{\overline{\text{MS}}}

\usepackage{color}

\definecolor{Brown}{rgb}{0.5,0.25,0}

\begin{document}


\begin{titlepage}

\begin{flushright}
 SI-HEP-2016-12\\
 QFET-2016-06\\[0.1cm]
 \today
\end{flushright}
\vskip 1.0cm

\begin{center}
\Large{\bf\boldmath
Two-body non-leptonic heavy-to-heavy \\ decays at NNLO in QCD factorization
\unboldmath}

\normalsize
\vskip 1.5cm

{\sc Tobias~Huber}$^{a}$, {\sc Susanne~Kr\"ankl}$^{a}$, {\sc Xin-Qiang~Li}$^{b}$\\

\vskip 0.8cm

{\it $^a$Naturwissenschaftlich-Technische Fakult\"at,
\\ Universit\"at Siegen, Walter-Flex-Str.\ 3, 57068 Siegen, Germany}\\[0.2cm]
{\it $^b$Institute of Particle Physics and \\ Key Laboratory of Quark and Lepton Physics~(MOE),
\\ Central China Normal University, Wuhan, Hubei 430079, P.\ R.\ China}

\vskip 1.8cm

\end{center}

\begin{abstract}

We evaluate in the framework of QCD factorization the two-loop vertex corrections to the decays $\bar{B}_{(s)}\to D_{(s)}^{(\ast)+} \, L^-$ and $\Lambda_b \to \Lambda_c^+ \, L^-$, where $L$ is a light meson from the set $\{\pi,\rho,K^{(\ast)},a_1\}$. These decays are paradigms of the QCD factorization approach since only the colour-allowed tree amplitude contributes at leading power. Hence they are sensitive to the size of power corrections once their leading-power perturbative expansion is under control. Here we compute the two-loop ${\cal O}(\alpha_s^2)$ correction to the leading-power hard scattering kernels, and give the results for the convoluted kernels almost completely analytically. Our newly computed contribution amounts to a positive shift of the magnitude of the tree amplitude by $\sim 2$\%. We then perform an extensive phenomenological analysis to NNLO in QCD factorization, using the most recent values for non-perturbative input parameters. Given the fact that the NNLO perturbative correction and updated values for form factors increase the theory prediction for branching ratios, while experimental central values have at the same time decreased, we reanalyze the role and potential size of power corrections by means of appropriately chosen ratios of decay channels.

\end{abstract}

\vfill

\end{titlepage}


\section{Introduction}

Non-leptonic two-body decays of bottom mesons and baryons are interesting for phenomenological studies of the quark flavour sector of the Standard Model~(SM) of particle physics. They yield observables like branching ratios and CP asymmetries that are relevant for studying the CKM mechanism of quark flavour mixing and allow access to the quantities of the unitarity triangle (cf.\ refs.~\cite{Buchalla:2008jp,Antonelli:2009ws,Aaij:2016oso}).

Oscillations and decays of $B$-mesons received considerable attention for the first time in the 1980s and 90s when the experiments ARGUS at DESY and CLEO at Cornell started to collect a lot of statistics. In the last decade, non-leptonic two-body $B_{(s)}$-decays have been extensively measured at the asymmetric $e^+e^-$ colliders ($B$-factories) at SLAC and KEK, but also in hadronic environments such as the Tevatron, and the results obtained by the Babar, Belle, D0 and CDF collaborations have reached a high level of precision (see, e.g.~\cite{Bevan:2014iga}). In recent years the LHCb experiment at the LHC at CERN has become the main player as far as experimental physics of the bottom quark is concerned. A large data set on bottom mesons and baryons has been accumulated, and results related to non-leptonic decays have been published (cf.~\cite{Aaij:2014jpa,Aaij:2015dsa}) and further analyses are ongoing. In the near future also Belle~II will contribute significantly to further improve the measurements~\cite{Aushev:2010bq}.

With the plethora of precise experimental data on non-leptonic decays at hand, theoretical predictions at the same level of accuracy are very much desired. However, the theoretical description of non-leptonic two-body $B_{(s)}$ decays is notoriously complicated. A straightforward computation of the hadronic matrix elements which describe the weak transition is not feasible due to the presence of the strong interaction in the purely hadronic initial and final states. This circumstance entails QCD effects from many different scales which are, moreover, largely separated. In a first approach, known as na\"ive factorization, the hadronic transition matrix elements were factorized into a product of a form factor and a decay constant~\cite{Bauer:1986bm}. Subsequent studies built on flavour symmetries of the light quarks~\cite{Zeppenfeld:1980ex} and on factorization frameworks such as perturbative QCD (pQCD)~\cite{Keum:2000ph,Keum:2000wi} and QCD factorization (QCDF)~\cite{Beneke:1999br,Beneke:2000ry,Beneke:2001ev}, to mention the most prominent ones. Certain combinations of these approaches can also be found (see e.g.~\cite{Zhou:2015jba}).

In the present work we adopt the QCDF framework and consider non-leptonic heavy-to-heavy transitions, which at the quark-level are mediated by the weak decay $b\to c \bar u d(s)$, where we treat the bottom and the charm quark as massive and the light quarks as massless. Performing an expansion of the amplitude in powers of $\Lamqcd/m_b$, where $\Lamqcd$ is the typical hadronic scale, a systematic separation of QCD effects from different scales can be achieved and corrections to na\"ive factorization be systematically included. Taking the decay $\bar{B}\to D^+ L^-$ as an example, the transition amplitude in the heavy-mass limit is then given by~\cite{Beneke:2000ry}
\begin{align}
 \langle D^{+}L^{-}|\mathcal{Q}_i |\bar{B} \rangle = \sum_j F_j^{B\rightarrow D} (m_L^2) \int_0^1 du \, T_{ij}(u) \Phi_{L}(u) \, , \label{bbnsfactorization}
\end{align}
where the local four-fermion operators $\mathcal{Q}_i $ describe the underlying weak decay. The $ F_j^{B\rightarrow D}$ form factors and the light-cone distribution amplitude (LCDA) $\Phi_{L}$ of the light meson contain long-distance effects and can be obtained from non-perturbative methods like
QCD sum rules and lattice QCD. The hard-scattering kernels $T_{ij}$, on the other hand, only receive contributions from scales of $\mathcal O(m_b)$ and are accessible in a perturbative expansion in the strong coupling $\al_s$. After the convolution over the momentum fraction $u$ of the valence quark inside the light meson, they yield a perturbative contribution to the topological tree amplitude $a_1(D^{+}L^{-})$. Taking the decay $\bar B \to D^+ \pi^-$ as a specific example, the latter is defined via~\cite{Beneke:2000ry}
\begin{align}
 {\mathcal{A}}(\bar B \to D^+ \pi^-) & = i \, \frac{G_F}{\sqrt{2}} \, V^\ast_{ud} \, V_{cb} \; a_1(D^+ \pi^-) \, f_\pi \, F_0^{B\to D}(m^2_\pi) \, (m_B^2-m_D^2) \, . \label{eq:defa1}
\end{align}

The leading-power hard-scattering kernels have been known to next-to-leading order (NLO) accuracy for more than a decade for both heavy-to-light~\cite{Beneke:1999br,Beneke:2001ev,Beneke:2003zv} and heavy-to-heavy~\cite{Beneke:2000ry} decays.
In the latter case, expanding the LCDA in Gegenbauer moments up to the first moment $\al^{L}_{1}$, the topological tree amplitude $a_1$ to NLO reads~\cite{Beneke:2000ry}
 \begin{align}
| a_1(\bar{B}\to D^+ L^-)|&=(1.055^{+0.019}_{-0.017})  -(0.013^{+0.011}_{-0.006}) \al^{L}_{1} \, , \nonumber \\
| a_1(\bar{B}\to D^{\ast+} L^-)|&=(1.054^{+0.018}_{-0.017})  -(0.015^{+0.013}_{-0.007}) \al^{L}_{1} \, . \label{a1NOL}
 \end{align}
For the light meson being $\pi$ or $\rho$ we have $\al^{\pi(\rho)}_{1}=0$ and for the kaon $|\al^{K}_{1}|<1$ is assumed~\cite{Beneke:2000ry}. With this mild dependence on the light meson LCDA we encounter a quasi-universal value $|a_1|\simeq 1.05$ for heavy-to-heavy decays in QCDF to NLO accuracy.
A quasi-universality was also found upon extracting $a_1$ from experimental data~\cite{Fleischer:2010ca}. However, the favoured central value $|a_1|\simeq 0.95$ for the decays $\bar{B}\to D^{(\ast)+} L^-$ ($L = \pi$, $K$), with errors in the individual channels at the $10$~--~$20$\% level, is considerably lower.

In recent years next-to-next-to-leading order (NNLO) corrections to heavy-to-light decays have become available~\cite{Li:2005wx,Beneke:2005vv,Li:2006jb,Kivel:2006xc,Beneke:2006mk,Jain:2007dy,Pilipp:2007mg,Bell:2007tv,Bell:2009nk,Bell:2009fm,Beneke:2009ek,Kim:2011jm,Bell:2014zya,Bell:2015koa}, and besides the prospects of increasing precision on the experimental side, there is multiple motivation to go beyond NLO in heavy-to-heavy transitions as well:
First, the NLO correction is small since it is proportional to a small Wilson coefficient and, in addition, is colour-suppressed. At NNLO the colour suppression gets lifted and the large Wilson coefficient re-enters, and therefore the NNLO correction could be comparable in size to the NLO term.
Moreover, it is interesting to see whether the quasi-universality of $a_1$ persists at NNLO.
At leading power the decays $\bar{B}_{(s)} \to D_{(s)}^{(\ast)+}L^-$ receive only vertex corrections to the colour-allowed tree topology. Interactions with the spectator quark as well as the weak annihilation topology are power-suppressed~\cite{Beneke:2000ry} and there are neither contributions from penguin operators nor is there a colour-suppressed tree topology. Therefore, a precise knowledge of the colour-allowed tree amplitude $a_1$ allows to reliably estimate the size of power corrections to eq.~\eqref{bbnsfactorization} by comparison to experimental data, and at the same time provides a test of the QCDF framework. This requires that the perturbative expansion of the hard scattering kernel is under control, and that also the uncertainties of the non-perturbative input parameters (form factors, decay constants, LCDAs) can be minimized. In the present work we therefore calculate the two-loop vertex correction to the leading-power hard scattering kernels in the framework of QCDF. Parts of the computational procedure were already presented in~\cite{Huber:2014kaa,Huber:2015bva}. Here, we give the full result of the technically challenging two-loop calculation. Besides, we present an updated phenomenological analysis of $\bar B_{(s)}\to D_{(s)}^{(*)+}L^-$ decays, with a light meson from the set $L=\{\pi,\rho,K^{(\ast)},a_1\}$\footnote{We use the same symbol $a_1$ for both, the meson $a_1(1260)$ and the colour-allowed tree amplitude $a_1(D^{+}L^{-})$. We think that in each case it is clear from the context which quantity we refer to.}, using the most recent values for non-perturbative input parameters (for another recent analysis, see~\cite{Chang:2016eto}).

Recently, non-leptonic $\Lambda_b$ decays have received considerable attention as well. Data on $\Lambda_b \to \Lambda_c^+ L^-$ with $L$ being $\pi$ or $K$~\cite{Aaij:2013pka} and on baryonic form factors have become available~\cite{Detmold:2015aaa}. Therefore, we extend our study to these decays.
Factorization has not yet been systematically established for baryonic decays, but was discussed in ref.~\cite{Leibovich:2003tw}. As a systematic derivation of the baryonic factorization formula is beyond the scope of this work we adopt the factorization formula eq.~(4) of ref.~\cite{Leibovich:2003tw}, with appropriate modifications to take perturbative corrections into account.

This article is organized as follows: In section~\ref{sec:theory} we present our theoretical framework by specifying our operator basis in the effective weak Hamiltonian. Subsequently, we derive the master formulas for the hard scattering kernels by performing a matching onto Soft-Collinear Effective Theory. In section~\ref{sec:technical} we discuss the calculation of the two-loop Feynman diagrams and specify the input to the master formulas. The analytical results of the hard scattering kernels after the convolution with the LCDAs are presented in section~\ref{sec:convoluted}. In section~\ref{sec:msbar} we give the formulas for converting from the pole to the $\overline{\text{MS}}$ scheme for the $b$- and $c$-quark masses. We present the results of our extensive phenomenological analysis in section~\ref{sec:pheno}, and conclude in section~\ref{sec:conclusion}.

\section{Theoretical framework}
\label{sec:theory}

\subsection{Five-flavour theory}
\label{sec:effectiveH}

We work in the effective five-flavour theory where the top quark, the heavy gauge bosons $W^\pm$, $Z^0$ and the Higgs boson are integrated out and their effects are absorbed into short-distance Wilson coefficients. The decays $\bar B_{(s)}\to D^{(*)+}_{(s)} \, L^-$ and $\Lambda_b \to \Lambda_c^+ \, L^-$ are mediated at parton level by a $b\to c \bar u d(s)$ transition for $L=\pi, \, \rho, \, a_1 \, (K, \, K^\ast)$. The corresponding QCD amplitude is computed in the framework of the effective weak Hamiltonian~\cite{Buchalla:1995vs,Beneke:2001ev}, which for the problem at hand simply reads
\begin{align}
 \mathcal H_\text{eff} = \frac{G_F}{\sqrt{2}}  V_{cb}V^*_{ud} \left(C_1 \mathcal Q_1 +C_2\mathcal Q_2  \right) + \text{h.c.} \, . \label{Hamiltonianfull}
\end{align}
We restrict our notation to the case of a $b\to c \bar u d$ transition. The expressions for a strange quark in the final state are obtained by obvious replacements. The local current-current operators in the Chetyrkin-Misiak-M\"unz (CMM) basis \cite{Chetyrkin:1996vx,Chetyrkin:1997gb} read
\begin{align}
 \mathcal{Q}_1&= \bar{c} \gam^{\mu} (1-\gam_5)T^A b  \hspace{2.5mm} \bar{d} \gam_{\mu} (1-\gam_5)T^A u  \label{Q1} \, , \\
\mathcal{Q}_2&=  \bar{c} \gam^{\mu} (1-\gam_5) b \hspace{2.5mm}  \bar{d} \gam_{\mu} (1-\gam_5)u  \, , \label{Q2}
 \end{align}
where $\mathcal{Q}_1$ and $\mathcal{Q}_2$ are referred to as colour-octet and colour-singlet operator, respectively.
 The use of the CMM basis allows for a consistent treatment of $\gam_5$ in the na\"ive dimensional regularization scheme with fully anti-commuting $\gam_5$.

Moreover, as the computation will be performed in dimensional regularization, we have to augment our physical operators $\mathcal Q_{1,2}$ by a set of evanescent operators, for which we adopt the convention~\cite{Gambino:2003zm,Gorbahn:2004my}
\begin{align}
 E_1^{(1)}&=   \left[\bar{c} \gam^{\mu}\gam^{\nu}\gam^{\rho} (1-\gam_5)T^A b \right]  \left[\bar{d} \gam_{\mu}\gam_{\nu}\gam_{\rho} (1-\gam_5)T^A u \right]- 16\mathcal Q_1 \, , \label{eq:E11}\\
 E_2^{(1)}&=    \left[\bar{c} \gam^{\mu}\gam^{\nu}\gam^{\rho} (1-\gam_5) b\right]  \left[\bar{d} \gam_{\mu}\gam_{\nu}\gam_{\rho} (1-\gam_5) u\right]- 16\mathcal Q_2 \, , \label{eq:E21}\\
  E_1^{(2)}&=    \left[\bar{c} \gam^{\mu}\gam^{\nu}\gam^{\rho}\gam^{\sig}\gam^{\lambda} (1-\gam_5)T^A b \right]  \left[\bar{d} \gam_{\mu}\gam_{\nu}\gam_{\rho}\gam_{\sig}\gam_{\lambda} (1-\gam_5)T^A u\right] -20 E_1^{(1)} - 256\mathcal Q_1 \, ,\\
    E_2^{(2)}&=  \left[\bar{c} \gam^{\mu}\gam^{\nu}\gam^{\rho}\gam^{\sig}\gam^{\lambda} (1-\gam_5)b\right] \left[ \bar{d} \gam_{\mu}\gam_{\nu}\gam_{\rho}\gam_{\sig}\gam_{\lambda} (1-\gam_5)u\right]  -20 E_2^{(1)} - 256\mathcal Q_2  \label{eq:E22}\,.
 \end{align}
These unphysical operators vanish in $D=4$ dimensions but contribute if $D\neq 4$ since they mix under renormalization with the physical operators. At two-loop accuracy the set of operators~\eqref{Q1}~--~\eqref{eq:E22} closes under renormalization.

\subsection{Matching onto SCET and master formulas}
\label{sec:matching}

We construct the master formulas for the hard scattering kernels by performing a matching from the effective weak Hamiltonian\footnote{In the following we refer to this side of the matching equation as the QCD side.} onto Soft-Collinear Effective Theory (SCET) with three light flavours. The procedure follows similar lines than the derivation of the master formulas for the hard kernels in heavy-to-light transitions~\cite{Beneke:2009ek}.

\begin{figure}[t]
\centering
 \includegraphics[width=6cm]{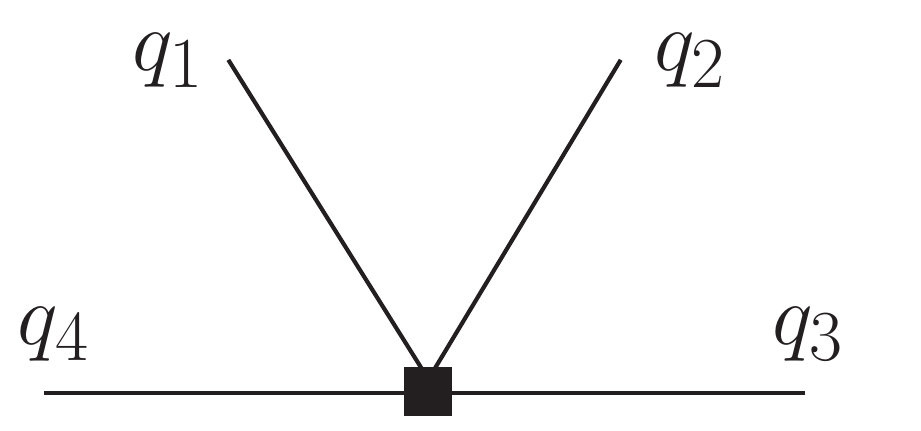} \vspace{0.3mm}
\caption{The tree-level Feynman diagram for the $b\to c \bar ud$ transition in full (five flavour) QCD: the black square represents the vertex of the effective weak interaction. The momenta $q_4$ and $q_3$ belong to the quark lines with masses $m_b$ and $m_c$, respectively, and $q_1+q_2 =q$ is the momentum of the light meson. All momenta are taken to be incoming. \label{fig:kinematics} }
\end{figure}

The kinematics of the $b\to c\bar ud$ transition is shown in the tree-level Feynman diagram depicted in figure~\ref{fig:kinematics}.
The $b$ and the $c$ quark are considered to be massive and carry momenta $q_4$ and $q_3$, respectively.
The massless $d$ and $\bar u$ quarks share the momentum $q$ with $q_1= u q$ and $q_2 =(1-u) q\equiv \bar u q$, where  $u\in [0,1]$ is the momentum fraction of the valence quark inside the light meson. All external momenta are taken to be incoming and are subject to the on-shell constraints $q_{4}^2=m_{b}^2$, $q_{3}^2=m_{c}^2$, and $q^2=0$.

We consider a reference frame in which the $b$ quark within the $B$ meson moves with momentum $q_b = m_b v+ k$, where $k$ is a residual momentum of order of the typical hadronic scale $\Lamqcd$, and $v$ is the velocity of the $B$ meson. The $b$ quark can then be described by a heavy-quark field $h_v$ which satisfies the equation of motion $\slashed v h_v= h_v$. We further choose a reference frame such that the energetic light meson moves in the light-cone direction $n_+$. The light-like vectors $n_+$ and $n_-= 2v-n_+$ then fulfill the constraints $n_\pm^2=0$ and $n_+ n_- =2$. As the quark and the anti-quark in the light meson nearly move in the same direction we can describe them by the same type of collinear SCET field $\chi$, which satisfies the equations of motion $\slashed n_+\chi =0$ and $\bar{\chi} \slashed n_+=0$.
In the derivation of the factorization formula~\eqref{bbnsfactorization} the power counting $ m_c/m_b\sim {\cal O}(1)$ was adopted. Hence, we treat the charm quark as a heavy quark and consequently describe it -- in analogy to the $b$ quark -- by another heavy-quark field $h_{v'}$ with velocity $v'$ and equation of motion $\slashed v' h_{v'}= h_{v'}$.

The amplitudes in full QCD and in SCET are made equal by adjusting the corresponding hard coefficients at the matching scale. We express the renormalized matrix elements of the QCD operators~\eqref{Q1} and~\eqref{Q2} as a linear combination of a basis of SCET operators,
\begin{align}
\langle \mathcal{Q}_i \rangle = \sum_{a=1}^3 \left[ H_{ia} \langle \mathcal{O}_a \rangle + H'_{ia} \langle \mathcal{O}'_a \rangle \right] \, , \label{matchingansatz}
\end{align}
where $H_{ia}$ and $ H'_{ia}$ are the matching coefficients. The basis of SCET operators is given by
\begin{align}
 \mathcal{O}_1= & \bar{\chi} \frac{ \slashed{n}_{-}}{2} (1-\gam_5) \chi \hspace{2.5mm} \bar{h}_{v'} \slashed{n}_{+} (1-\gam_5) h_v \, , \label{Op1}\\
\mathcal{O}_2= & \bar{\chi} \frac{ \slashed{n}_{-}}{2} (1-\gam_5) \gam_{\bot}^{\al} \gam_{\bot}^{\bet} \chi\hspace{2.5mm} \bar{h}_{v'} \slashed{n}_{+} (1-\gam_5) \gam_{\bot\bet} \gam_{\bot\al} h_v \,, \label{Op2QCD} \\
\mathcal{O}_3= & \bar{\chi} \frac{ \slashed{n}_{-}}{2} (1-\gam_5) \gam_{\bot}^{\al} \gam_{\bot}^{\bet}\gam_{\bot}^{\gam}\gam_{\bot}^{\del} \chi \hspace{2.5mm}  \bar{h}_{v'} \slashed{n}_{+} (1-\gam_5) \gam_{\bot\del} \gam_{\bot\gam}\gam_{\bot\bet} \gam_{\bot\al} h_v \, ,  \label{Op3} \\
 \mathcal{O}'_1= & \bar{\chi} \frac{ \slashed{n}_{-}}{2} (1-\gam_5) \chi \hspace{2.5mm}\bar{h}_{v'}\slashed{n}_{+} (1+\gam_5) h_v \,,  \label{Op1bar}\\
\mathcal{O}'_2= & \bar{\chi} \frac{ \slashed{n}_{-}}{2} (1-\gam_5) \gam_{\bot}^{\al} \gam_{\bot}^{\bet}  \chi \hspace{2.5mm} \bar{h}_{v'} \slashed{n}_{+} (1+\gam_5)\gam_{\bot\al} \gam_{\bot\bet}  h_v \, ,\\
\mathcal{O}'_3= & \bar{\chi} \frac{ \slashed{n}_{-}}{2} (1-\gam_5) \gam_{\bot}^{\al} \gam_{\bot}^{\bet}\gam_{\bot}^{\gam}\gam_{\bot}^{\del} \chi \hspace{2.5mm} \bar{h}_{v'} \slashed{n}_{+} (1+\gam_5)  \gam_{\bot\al}  \gam_{\bot\bet}  \gam_{\bot\gam} \gam_{\bot\del}h_v \, . \label{Op3bar}
\end{align}
Here, the perpendicular component of a Dirac matrix is defined by
\begin{align}
\gamma^{\mu} & = \slashed{n}_{+}  \, \frac{n_{-}^\mu}{2}+ \slashed{n}_{-} \,  \frac{n_{+}^\mu}{2} + \gamma_{\perp}^\mu .
\end{align}
Moreover, we have omitted the Wilson lines which render the non-local light currents $\bar{\chi} (t n_{-})[\dots] \chi (0)$ gauge invariant. One therefore has to keep in mind that the coefficients $H_{ia}$ are also functions of the variable $t$, and the products $H^{(\prime)}_{ia} \langle \mathcal{O}^{(\prime)}_a \rangle$ in eq.~\eqref{matchingansatz} are in fact convolutions. We also remark that the SCET operator basis is chosen such that all operators with index $a>1$ are evanescent, and we have the two physical SCET operators $\mathcal{O}_1$ and $\mathcal{O}'_1$. The operators~\eqref{Op1}~--~\eqref{Op3} have the same structure as those in~\cite{Beneke:2009ek} for heavy-to-light transitions, but with a heavy-quark field $h_{v'}$ instead of an anti-collinear SCET field $\xi$ in direction $n_-$. For heavy-to-heavy transitions this set of operators has to be extended by those in~\eqref{Op1bar}~--~\eqref{Op3bar} which have a different chirality structure, to take into account the non-vanishing mass of the charm quark. For technical details on the operators see~\cite{Neubert:1993mb,Beneke:2005vv}.

We first consider the expansion of the left-hand side of eq.~\eqref{matchingansatz} in terms of on-shell QCD amplitudes.
The expression for the renormalized matrix elements reads
\begin{align}	
 \langle \mathcal{Q}_i\rangle& = \bigg\{ A_{ia}^{(0)} +\alsfpi \left[ A_{ia}^{(1)}+ Z_{ext}^{(1)} A_{ia}^{(0)}+Z_{ij}^{(1)} A_{ja}^{(0)}\right] \nonumber \\
&\quad + \left(\alsfpi \right)^2 \Big[ A_{ia}^{(2)}+  Z_{ij}^{(1)} A_{ja}^{(1)}  + Z_{ij}^{(2)} A_{ja}^{(0)}+ Z_{ext}^{(1)} A_{ia}^{(1)}+ Z_{ext}^{(2)} A_{ia}^{(0)} +Z_{ext}^{(1)} Z_{ij}^{(1)} A_{ja}^{(0)}\nonumber \\
&\quad +(-i)\del m_b^{(1)} {A}_{ia}^{*(1)} + (-i)\del m_c^{(1)} A_{ia}^{**(1)} +Z_\al^{(1)} A_{ia}^{(1)}  \Big] +\mathcal{O}\big(\al_s^3\big) \bigg\} \langle \mathcal{O}_a	\rangle^{(0)}  \nn \\
& \quad + ( A \leftrightarrow A') \langle \mathcal{O}_a'\rangle^{(0)}
\label{QCDside} \, .
\end{align}
Here, a sum over $a=1,2,3$ is understood, and $\al_s$ is the $\overline{\text{MS}}$ strong coupling constant
with five active flavours. The index $i=1,2$ denotes the physical operators from \eqref{Q1} and \eqref{Q2} only, whereas $j$ includes physical as well as evanescent operators from \eqref{Q1}~--~\eqref{eq:E22}, hence $j=1,\dots,6$.
The $ A^{(l)}$  are the bare $l$-loop on-shell amplitudes and $ {A}^{*(1)}$ (${A}^{**(1)}$) is the one-loop bare on-shell amplitude with a $b$ ($c$) quark mass insertion on the heavy $b$ ($c$) line. The primed amplitudes are defined analogously.
The renormalization factors $Z_{ij}$, $Z_{ext}$ and $Z_{\al}$ stem from operator renormalization, wave-function renormalization of all external legs and coupling renormalization, respectively. They are defined in a perturbative expansion
\begin{align}
 Z = 1 +\sum_{k=1}^\infty \left(\frac{\al_s}{4\pi}\right)^{\! k} \, Z^{(k)} \, . \label{eq:Z}
\end{align}
The operator renormalization is performed in the $\overline{\text{MS}}$ scheme, whereas for the
mass and the wave-function renormalization the on-shell scheme is applied.
Renormalized matrix elements of evanescent operators vanish also beyond tree level. Nevertheless, these operators
cannot be neglected right from the beginning as they yield physical contributions to the products $Z_{ij}^{(1)} A_{ja}^{(0)}$, $Z_{ij}^{(1)} A_{ja}^{(1)}$, and $Z_{ij}^{(2)} A_{ja}^{(0)}$.

Similarly, we can write down the expression for the renormalized matrix elements of the SCET operators that enter the right-hand side of eq.~\eqref{matchingansatz} and obtain
\begin{align}
 \langle \mathcal{O}_a \rangle = \bigg\{ \del_{ab} &+   \alsfpihat \left[ M_{ab}^{(1)} + Y_{ext}^{(1)} \del_{ab} + Y_{ab}^{(1)} \right] \, \nonumber \\
&+ \left( \alsfpihat \right)^2\Big[ M_{ab}^{(2)} +Y_{ext}^{(1)} M_{ab}^{(1)} +Y_{ac}^{(1)} M_{cb}^{(1)}+ \hat{Z}_{\al}^{(1)} M_{ab}^{(1)}+ Y_{ext}^{(2)} \del_{ab} \nonumber \\
& +Y_{ext}^{(1)} Y_{ab}^{(1)}+ Y_{ab}^{(2)}  \Big]  + \mathcal{O} \big(\alshat^3\big)   \bigg\} \langle \mathcal{O}_b\rangle^{(0)}
\,. \label{SCETside}
\end{align}
Here, $a=1,2,3$ and a sum over $b=1,2,3$ is understood. The  $\overline{\text{MS}}$ strong coupling constant $\alshat$ has three  light flavours and $M^{(l)}$ are the bare $l$-loop SCET amplitudes.
The $Y_{ext}^{(l)} $, $ Y_{ab}^{(l)}$ and $\hat Z_{\al}^{(l)}$ are the $l$-loop wave-function, operator and coupling renormalization constants, respectively. They are defined in a perturbative expansion analogous to eq.~\eqref{eq:Z} except that the strong coupling has only three light flavours. The corresponding expression for the primed operators from eqs.~\eqref{Op1bar}~--~\eqref{Op3bar} is given by substituting $M\rightarrow M'$ and
 $\mathcal O \rightarrow \mathcal O '$ in eq.~\eqref{SCETside}.

Eq.~\eqref{SCETside} can be simplified to a large extent.
In dimensional regularization
the on-shell renormalization constants $Y_{ext}$ are equal to unity. Moreover, the bare on-shell amplitudes only contain scaleless integrals, which vanish in dimensional regularization. We thus arrive at the following simplified expression of eq.~\eqref{SCETside}
\begin{align}
 \langle \mathcal{O}_a \rangle = \bigg\{ \del_{ab} &+  \alsfpihat Y_{ab}^{(1)}  + \left(\alsfpihat\right)^2   Y_{ab}^{(2)} + \mathcal{O}\left(\alshat^3\right)\bigg\} \langle \mathcal{O}_b\rangle^{(0)} \,,	 \label{SCETside2}
\end{align}
which for the primed operators takes a similar form.

For relating the matching coefficients $H_{ia}$ and $H_{ia}^{\prime}$ in eq.~\eqref{matchingansatz} to the hard scattering kernels  we introduce two factorized QCD operators
\begin{align}
 \mathcal{Q}^{(')\text{QCD}}= \displaystyle [ \bar{q} \frac{\slashed{n}_{-}}{2} (1-\gam_5) q ] [ \bar{c} \, \slashed{n}_{+} (1 \mp\gam_5) b ] \,  ,
\end{align}
which are by definition the products of the two currents in brackets. The upper sign corresponds to the un-primed operator. The renormalized operators $\mathcal{Q}^{(')\text{QCD}}$ are then matched onto the renormalized SCET operators $\mathcal{O}_1$ and $\mathcal{O}_1'$ by adjusting the corresponding hard coefficients. This can be done separately for the light-to-light and heavy-to-heavy currents. For the renormalized light-to-light current we make the ansatz
\begin{align}
 \bigg[\bar{q} \frac{\slashed{n}_{-}}{2} (1-\gam_5) q \bigg] &= C_{\bar{q}q} \bigg[\bar{\chi} \frac{\slashed{n}_{-}}{2} (1-\gam_5) \chi\bigg] \, . \label{matchcurrentsl}
\end{align}
The heavy-to-heavy currents with different chiralities mix in the matching. Thus, we make the ansatz
\begin{align}
 \big[\bar{c} \, \slashed{n}_{+} (1-\gam_5) b\big]\;&= C_{FF}^\text{LL} \, \big[ \bar{h}_{v'}   \slashed{n}_{+} (1-\gam_5) h_v \big] + C_{FF}^\text{LR} \, \big[ \bar{h}_{v'}   \slashed{n}_{+} (1+\gam_5) h_v \big]\,,  \\
\big[\bar{c} \, \slashed{n}_{+} (1+\gam_5) b\big] \; &=C_{FF}^\text{RL} \, \big[ \bar{h}_{v'}   \slashed{n}_{+} (1-\gam_5) h_v \big] + C_{FF}^\text{RR} \, \big[ \bar{h}_{v'}   \slashed{n}_{+} (1+\gam_5) h_v \big]\, . \label{matchcurrentsh}
\end{align}
Since these equations are symmetric under interchanging $P_L \leftrightarrow P_R$ we have $C_{FF}^{\text{LL}} =  C_{FF}^\text{RR} \equiv C_{FF}^{\text{D}} $ and $C_{FF}^\text{LR} =C_{FF}^\text{RL} \equiv C_{FF}^{\text{ND}}$.
Finally, we obtain
\begin{align}
 \mathcal{Q}^{\text{QCD}}= \bigg[ \bar{q} \frac{\slashed{n}_{-}}{2} (1-\gam_5) q  \bigg] \big[ \bar{c} \, \slashed{n}_{+} (1-\gam_5) b \big] &=  C_{\bar{q}q} C_{FF}^\text{D} \mathcal{O}_1 +C_{\bar{q}q} C_{FF}^\text{ND} \mathcal{O}_1' \label{factorizedQCD}   \, ,\\
 \mathcal{Q}^{'\text{QCD}}=\bigg[ \bar{q} \frac{\slashed{n}_{-}}{2} (1-\gam_5) q  \bigg] \big[ \bar{c} \, \slashed{n}_{+} (1+\gam_5) b \big] &=  C_{\bar{q}q}C_{FF}^\text{ND} \mathcal{O}_1 +C_{\bar{q}q}C_{FF}^\text{D} \mathcal{O}_1' \, . \label{factorizedQCDpr}
\end{align}
Since by construction $ \mathcal{Q}^{\text{QCD}}$ and  $\mathcal{Q}^{'\text{QCD}}$
factorize into a light-to-light and a heavy-to-heavy current, the matrix element of these operators
is the product of an LCDA and the full QCD form factor with the corresponding helicity structure.

We now consider the two hard scattering kernels $\hat T_{i}$ and  $\hat T'_{i}$ that are defined by the following expression
\begin{align}
\langle \mathcal{Q}_i \rangle =  \hat T_{i} \langle \mathcal{Q}^{\text{QCD}} \rangle + \hat T_{i}' \langle \mathcal{Q}^{'\text{QCD}} \rangle +\sum_{a>1} \left[ H_{ia} \langle \mathcal{O}_a \rangle +  H'_{ia} \langle \mathcal{O}'_a \rangle \right] \, .  \label{matchingansatzQCD}
\end{align}
Comparing eqs.~\eqref{matchingansatz} and~\eqref{matchingansatzQCD}, $\hat T_{i}$ and  $\hat T'_{i}$ can be related to the matching coefficients as follows
\begin{align}
 \begin{pmatrix}
  \hat T_i  \\ \hat T_i'
 \end{pmatrix}
 = \begin{pmatrix}
    C_{\bar q q} C_{FF}^\text{D} &	 C_{\bar q q} C_{FF}^\text{ND}  \\
     C_{\bar q q} C_{FF}^\text{ND} &  C_{\bar q q} C_{FF}^\text{D}
   \end{pmatrix}^{-1}
   \begin{pmatrix}
    H_{i1} \\ H_{i1}'
   \end{pmatrix} \, .
\label{Tgeneral}
\end{align}
Plugging in the  matching coefficients as expansions in the five-flavour coupling $\al_s$, the matrix can be inverted order-by-order in $\al_s$.
We remark that $C_{\bar{q}q}=1+\mathcal O(\al_s^2)$, i.e.\ it receives a correction at two loops only since at one loop only scaleless integrals contribute. The explicit one-loop expressions for the heavy-to-heavy coefficients will be derived in section~\ref{sec:technical}.  For the diagonal coefficients we have $ C_{FF}^\text{D}= 1+\mathcal O(\al_s)$. In contrast, the non-diagonal matching coefficients  $ C_{FF}^\text{ND}$ that induce the chirality mixing only arise beyond tree level, $ C_{FF}^\text{ND}= \mathcal O(\al_s)$.

Putting everything together, the master formulas for the hard scattering kernels read
\begin{align}
  \hat {T}_i^{(0)} &= A_{i1}^{(0)}  \nn \\
 \hat {T}_i^{(1)}  &=  A_{i1}^{(1)nf}  +Z_{ij}^{(1)} A_{j1}^{(0)} \nn \\
 \hat  T_i^{(2)} &= A_{i1}^{(2)nf}+ Z_{ij}^{(1)}A_{j1}^{(1)} +Z_{ij}^{(2)}A_{j1}^{(0)}    + Z_\al^{(1)} A_{i1}^{(1)nf}  -\hat T_i^{(1)}  \left[  C_{FF}^{\text{D}(1)} + Y_{11}^{(1)} - Z_{ext}^{(1)}\right]\nn \\[1mm] &\quad
  - C_{FF}^{\text{ND}(1)}\hat T'^{(1)}_i  + (-i )\del m_b^{(1)} A_{i1}^{*(1)nf} + (-i )\del m_c^{(1)} A_{i1}^{**(1)nf}   -\sum_{b\neq 1}  H_{ib}^{(1)} Y_{b1}^{(1)}   \, . \label{T2octet}
\end{align}
The expression for the primed kernels $\hat  T'_i$ is given by eq.~\eqref{T2octet} with the replacement $A \leftrightarrow A',\, H\leftrightarrow H',\, \hat T\leftrightarrow \hat T'$. Note that the quantities $H^{(l)}$, $A^{(l)}$ and the hard kernels $\hat T^{(l)}$ depend on the quark mass ratio $z_c=m_c^2/m_b^2$ and the momentum fraction $u$ of the quark inside the light meson (as do the correspon\-ding primed quantities). Whenever they appear alongside a renormalization factor $Y^{(l)}$ such as $H_{ib}^{(1)} Y_{b1}^{(1)}$ we must keep in mind that these expressions must be interpreted as a convolution product $\int_0^1 du'\, H_{ib}^{(1)}(z_c,u')Y_{b1}^{(1)}(u',u)$.

\begin{figure}[t]
  \begin{minipage}{0.3\textwidth}
 \includegraphics[width=1.0\textwidth]{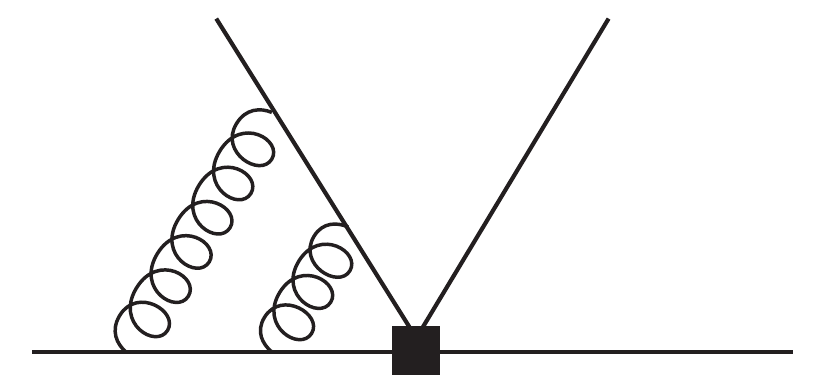} \vspace{0.3mm}
\end{minipage} \hspace{2mm}
  \begin{minipage}{0.3\textwidth}
\includegraphics[width=1.0\textwidth]{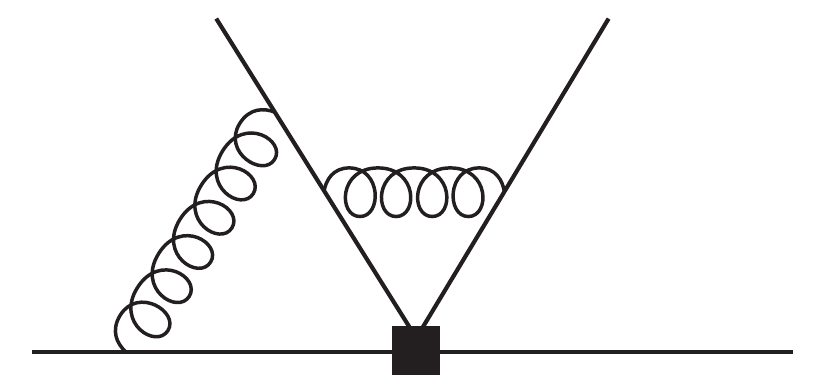} \vspace{0.3mm}
\end{minipage} \hspace{2mm}
  \begin{minipage}{0.3\textwidth}
\includegraphics[width=1.0\textwidth]{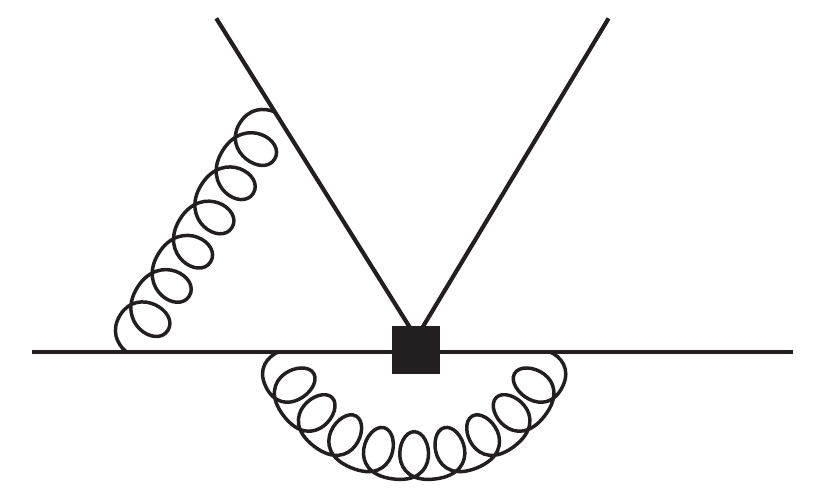}
\end{minipage}
\caption{Sample of Feynman diagrams that contribute to the two-loop hard scattering kernels.
\label{fig:sample} }
\end{figure}
 The amplitudes $A_{i1}^{(l)nf}$  in eq.~\eqref{T2octet} are termed ``non-factorizable''. At one-loop the corresponding amplitudes are given by all Feynman diagrams with one gluon connecting the heavy and the light current. The one-loop Feynman diagrams where the gluon is attached solely to either the light or the heavy current are part of the LCDA and the form factor, respectively.
 The Feynman diagrams contributing to $A_{i1}^{(2) nf}$ can be found in figures~15 and 16 of ref.~\cite{Beneke:2000ry} and in addition include the one-loop self-energy insertions to the ``non-factorizable'' one-loop amplitudes. A sample of two-loop diagrams is shown in figure~\ref{fig:sample}.
$A_{i1}^{(2)nf}$ is technically the most challenging contribution to the two-loop kernels. Therefore, we briefly describe their evaluation in the next section and, moreover, specify the remaining input to eq.~\eqref{T2octet}. The final expression of the hard scattering kernels must be free of ultraviolet and infrared divergences. We comment on this at the end of the next section.

Finally, we remark that eq.~\eqref{T2octet} has a structure similar to the corresponding expressions for the two-loop hard scattering kernel in the right-insertion contribution to the decay $B\to \pi \pi$, which is given in eq.~(24) in~\cite{Beneke:2009ek}. The main difference is three-fold: First, we find two contributions $\hat T$ and $\hat T'$ to the hard scattering kernel as a result of the extended operator basis. Second, we encounter the off-diagonal element $C_{FF}^{\text{ND}(1)}\hat T'^{(1)}_1$ due to the mixing of the heavy-to-heavy currents with different chirality structures. Finally, we have a mass counterterm for the massive charm quark in eq.~\eqref{T2octet}.

\section{Computational details}
\label{sec:technical}

\subsection{Technical aspects of the two-loop computation}
\label{sec:techtwoloops}

We work in dimensional regularization with $D=4-2\eps$ and expand the amplitudes in the parameter $\eps$.
The Feynman diagrams contributing to the bare two-loop amplitude $A^{(2)nf}$ then contain up to $1/\eps^4$ poles stemming from ultraviolet~(UV) and infrared~(IR) regions. We calculated them by applying commonly-used multi-loop techniques, including a new method for evaluating the master integrals. The procedure goes as follows: First, we decompose all tensor integrals into scalar ones by applying the Passarino-Veltman decomposition~\cite{Passarino:1978jh}.
We then perform the reduction of the Dirac structures to the SCET operator basis given in eqs.~\eqref{Op1}~--~\eqref{Op3bar} in {\tt Mathematica} by using simple algebraic transformations. The number of remaining scalar two-loop integrals exceeds several thousands and can be simplified by using the Laporta algorithm~\cite{Laporta:1996mq,Laporta:2001dd}, which is based on integration-by-parts identities~\cite{Chetyrkin:1981qh}.
Here, we apply the implementations AIR~\cite{Anastasiou:2004vj} (in {\tt Maple}) and FIRE~\cite{Smirnov:2008iw} (in {\tt Mathematica}) of this algorithm to reduce the large number of integrals to a small set of master integrals.
Many of the latter are already known from several $B\rightarrow \pi\pi$ calculations~\cite{Bell:2007tv,Bell:2009nk,Beneke:2009ek}. In addition, we find 23 yet unknown two-loop master integrals. Since most of them depend on two scales (the momentum fraction $u$ and the quark-mass ratio $z_c=m_c^2/m_b^2$), an analytic solution by common techniques is hardly feasible. We therefore evaluate them by applying the approach of differential equations in a canonical basis recently advocated in~\cite{Henn:2013pwa}. The solution is given by iterated integrals and falls into the class of Goncharov polylogarithms~\cite{Goncharov:1998kja}. We obtain analytic results for all 23 master integrals. Details on their calculation and the result of all master integrals can be found in~\cite{Huber:2015bva}.

\subsection{Input to the master formulas}
\label{sec:inputmaster}

Here we give the explicit expressions for the renormalization factors and matching coefficients that enter the master formula, and in the end comment on the cancellation of the poles in $\eps$ once all pieces of the master formula are plugged in.

The operator renormalization factors $Z_{ij}$ of the effective weak Hamiltonian were calculated to two-loop accuracy in the $\overline{\text{MS}}$ scheme in~\cite{Gambino:2003zm,Gorbahn:2004my}. The explicit one- and two-loop expressions read
\begin{align}
Z^{(1)}&=  \frac{1}{\eps} \left(
\begin{array}{cccccc}
 -2 & \frac{4}{3} & \frac{5}{12} & \frac{2}{9} & 0 & 0 \\
 6 & 0 & 1 & 0 & 0 & 0 \\
\end{array}
\right)
 \, ,  \label{Z1}\\[4mm]
Z^{(2)} &= \frac{1}{\eps^2}\left(
\begin{array}{cccccc}
 17-\frac{4 n_f T_f}{3} & \frac{2}{9} (4 n_f T_f-39) & \frac{5}{18} (n_f T_f-15) & \frac{1}{54} (8
   n_f T_f-93) & \frac{19}{96} & \frac{5}{108} \\[1mm]
 4 n_f T_f-39 & 4 & \frac{2 n_f T_f}{3}-\frac{31}{4} & 0 & \frac{5}{24} & \frac{1}{9} \\
\end{array}
\right) \nn \\[3mm]
&\quad +\frac{1}{\eps}\left(
\begin{array}{cccccc}
 \frac{8 n_f T_f}{9}+\frac{79}{12} & \frac{20 n_f T_f}{27}-\frac{205}{18} & \frac{1531}{288}-\frac{5 n_f
   T_f}{108} & -\frac{2 n_f T_f}{81}-\frac{1}{72} & \frac{1}{384} & -\frac{35}{864} \\[1mm]
 \frac{10 n_f T_f}{3}+\frac{83}{4} & 3 & \frac{119}{16}-\frac{n_f T_f}{9} & \frac{8}{9} & -\frac{35}{192} &
   -\frac{7}{72} \\
\end{array}
\right)
\, .
\end{align}
Here, $n_f=5$ is the total number of active quark flavours and $T_f=1/2$.
The row index of these matrices corresponds to $(\mathcal Q_1, \mathcal Q_2,E_1^{(1)},E_2^{(1)},E_1^{(2)},E_2^{(2)})$  and the column index to $(\mathcal Q_1,\mathcal Q_2)$. The strong coupling constant is renormalized in the $\overline{\text{MS}}$ scheme as well, whereas the renormalization of the masses and the wave-functions is performed in the on-shell scheme. The corresponding renormalization factors are well known and shall not be repeated here.

In eq.~\eqref{T2octet} we further encounter the SCET operator renormalization factor $Y_{11}$ that can be split into the following two parts
\begin{align}
Y_{11}(u',u)=Z_{Jh}\delta(u-u') +Z_{BL}(u',u) \, . \label{oprenscet}
\end{align}
Here, $Z_{Jh}$ and $Z_{BL}$ are the renormalization factors for the HQET heavy-to-heavy and the SCET light-to-light current, respectively. Since one collinear sector in SCET is equivalent to full QCD, the renormalization constant $Z_{BL}$ coincides with the ERBL kernel in QCD~\cite{Efremov:1979qk,Lepage:1980fj}.
We take $Z_{BL}$ from~\cite{Beneke:2005gs}, which for pseudoscalar and longitudinally polarized vector mesons reads
 \begin{align}
 Z_{BL}(v,w) = \del (v-w)   - \frac{\al_s }{4 \pi} &\frac{2 C_F }{ \eps}  \Bigg\{ \frac{1}{w \bar w}\left[v \bar w \,\frac{\Theta (w-v)}{w-v} +w \bar v \,\frac{\Theta (v-w)}{v-w} \right]_+
 -\frac{1}{2} \del(v-w)
 \nn \\
\quad & + \left[\frac{v}{w} \Theta(w-v)+\frac{\bar v}{\bar w} \Theta(v-w) \right]\Bigg\} + \mathcal O(\al_s^2) \, . \label{eq:ZBLdef}
\end{align}
The plus-distribution for symmetric kernels $f$ is defined as follows,
\begin{align}	
 \int dw \left[ f(v,w)\right]_+ g(w) = \int dw f(v,w) \, \left[g(w)-g(v)\right] \, .
\end{align}

The renormalization factor $Z_{Jh}$ can be obtained in a matching of the heavy-to-heavy QCD current $\bar c \,\slashed n_+ (1-\gam_5) b$ to the HQET current $\bar  h_{v'}  \slashed n_+ (1-\gam_5) h_v$. In this process also the matching coefficients $C_{FF}$ can be determined.
 Beyond tree-level the QCD current also mixes into the chirality-flipped HQET current $\bar  h_{v'}  \slashed n_+ (1+\gam_5) h_v$. Hence, we make the following ansatz for the renormalized currents
\begin{align}
 \bar{c} \, \slashed{n}_{+} (1\mp\gam_5) b  &= C_{FF}^\text{D} \left[ \bar  h_{v'}  \slashed n_+ (1\mp\gam_5) h_v \right]+ C_{FF}^\text{ND} \left[\bar  h_{v'}  \slashed n_+ (1\pm\gam_5) h_v \right]\, , \label{eq:CFF1}
\end{align}
where we have already made  use of the fact that both equations are symmetric under interchanging $P_L \leftrightarrow P_R$. The renormalization factor  $Z_{Jh}$ is defined via the on-shell  one-loop matrix element of the HQET currents
 \begin{align}
  \langle \bar  h_{v'} \,\slashed n_+ (1 \mp \gam_5)   h_{v}\rangle^{(1)} = \left(Y_{ext}^{(1)}+ Z_{Jh}^{(1)} \right)\langle \bar  h_{v'} \,\slashed n_+ (1 \mp\gam_5)   h_{v}\rangle^{(0)}\, . \label{eq:CFFscet}
 \end{align}
The one-loop renormalized matrix elements of the QCD currents can be calculated straightforwardly. Inserting their explicit expressions in eq.~\eqref{eq:CFF1} we can identify $Z_{Jh}^{(1)}$ as the pole term in $\eps$, that is
\begin{align}
 Z_{Jh}^{(1)} =  \frac{C_F }{\eps} \left(\frac{(z_c+1)  \log (z_c)}{z_c-1}-2\right)  \, ,
\end{align}
which correctly reproduces the IR behaviour of QCD currents in the effective theory.
The $C_{FF}$, on the other hand, are given by the coefficients that are finite in $\eps$. Their explicit
expressions read ($L\equiv \log (\mu^2/m_b^2)$)
\begin{align}
 C_{FF}^{\text{D}(1)} &=  C_F\left[L  \left(\frac{(z_c+1)
   \log (z_c)}{z_c-1}-2\right)  + \frac{(z_c+1) \log ^2(z_c)}{2-2 z_c}  + \frac{(5 z_c+1) \log(z_c)}{2 (z_c-1)}-4\right] \, ,\\
   C_{FF}^{\text{ND}(1)} &=C_F \left(\frac{\sqrt{z_c} \log (z_c)}{z_c-1} \right) \,.
\end{align}

As a last step the contribution of the sum $\sum_{b\neq 1}  H_{ib}^{(1)} Y_{b1}^{(1)}$ in eq.~\eqref{T2octet} needs to be further specified (the primed quantities are obtained by obvious substitutions).
We find that only $H_{ib}^{(1)}$ with $i=1$ and $b=2$ yields a non-vanishing contribution. A straightforward calculation yields $H_{12}^{(1)}= A_{12}^{(1)nf} +Z_{1j}^{(1)}A_{j2}^{(0)}$. The operator renormalization factor $Y_{21}^{(1)}$ has already been used in the NNLO calculation of the vertex corrections to the decay $B\to\pi\pi$ and is given in eq.~(45) of~\cite{Bell:2007tv}. Its explicit expression reads
\begin{align}
 Y_{21}^{(1)}(u',u)=16\, C_F\left(\frac{u' \Theta(u-u')}{u}+\frac{(1-u')\Theta (u'-u)}{1-u}\right) \, .
\end{align}
With this we have specified all input to the master formulas and are now ready to produce an expression for the hard scattering kernels.

The final expressions for the hard scattering kernels are free of poles in $\eps$, even though most of the individual terms in eq.~\eqref{T2octet} contain divergences. At the one-loop level we checked the cancellation of all poles analytically.
We find that our expressions for the finite pieces of the one-loop kernels agree with the results given in eqs.~(89) and~(90) in~\cite{Beneke:2000ry}\footnote{For performing this comparison one has to take into account that the one-loop result given in eq.~(90) in~\cite{Beneke:2000ry} was calculated in the \textquotedblleft traditional operator basis\textquotedblright\ given in eq.~(V.1) in~\cite{Buchalla:1995vs}.}. Some of the one-loop quantities that enter the two-loop master formula (last equation in~\eqref{T2octet}) have to be evaluated to higher orders in the $\eps$-expansion since they multiply poles in $\eps$ contained in the renormalization factors. We checked that in the limit $m_c \to 0$ the ${\mathcal O}(\eps)$ piece of the one-loop hard scattering kernel coincides with the one used in~\cite{Beneke:2009ek}.

At two loops we could check the pole cancellation numerically to an accuracy of $1 \times 10^{-10}$ or better for 12 different points in the $u$-$z_c$ plane. To this end, we evaluate the Goncharov polylogarithms and the harmonic polylogarithms~\cite{Remiddi:1999ew} that are contained in $A^{(2)nf}$ numerically with the C++ routine GiNaC~\cite{Vollinga:2004sn} and the {\tt Mathematica} program {\tt HPL}~\cite{Maitre:2005uu,Maitre:2007kp}, respectively. The explicit results for the two-loop hard scattering kernels are lengthy, not very illuminating, and enter the physical quantities only after convolution with the LCDAs. For these reasons we refrain from presenting them explicitly here, but they can be obtained from the authors upon request. However, after the convolution of the hard scattering kernels with an expansion of the LCDAs in terms of Gegenbauer polynomials up to the second moment, the expressions simplify considerably and we can express the result almost entirely in terms of harmonic polylogarithms. At this level we convolute also the pole terms in $\eps$ and checked that for the convoluted kernels all poles cancel analytically. We give the corresponding finite parts in the next section.

\section{Convoluted kernels}
\label{sec:convoluted}

The light meson LCDAs are expanded in a basis of Gegenbauer polynomials $C^{3/2}_k(x)$ with Gegenbauer moments $\al^L_k$,
\begin{align} 	
 \Phi_L(u,\mu)= 6 u (1-u) \left( 1+ \sum_{k=1}^{\infty} \al^L_k(\mu) C^{3/2}_k(2u-1)  \right) \, .\label{gegenbauer}
\end{align}
 Following~\cite{Beneke:2000ry} we assume that the leading-twist LCDA is close to its asymptotic form $ \Phi_L(u,\mu)= 6 u (1-u)$ and truncate the expansion after the second moment. The first two Gegenbauer polynomials read $C^{3/2}_1(x)= 3x $ and $C^{3/2}_2(x)= \frac{3}{2} (5 x^2-1)$.
The Gegenbauer polynomials are eigenfunctions of the one-loop renormalized ERBL-kernel~\cite{Mueller:1993hg} and thus the Gegenbauer moments are multiplicatively renormalizable to leading-logarithmic (LL) accuracy~\cite{Mueller:1993hg}.
 The next-to-leading logarithmic (NLL) evolution was derived in~\cite{Mueller:1993hg,Mikhailov:1984ii,Mueller:1994cn}.

The result for the hard scattering kernels after the convolution with the LCDA can be written as follows
\begin{align}
 \int_0^1 \!\! du \; \hat T_{i}(u,\mu) \Phi_L(u,\mu) &=  V_{i}^{(0)}(\mu) +\sum_{l\geq 1} \left(\frac{\al_s}{4\pi} \right)^{\! l \,\,}  \sum_{k=0}^{2} \al_k^L(\mu) \, V_{ik}^{(l)}(\mu) \, ,  \label{Tconv}\\
 \int_0^1 \!\! du \; \hat T'_{i}(u,\mu) \Phi_L(u,\mu) & =  V'^{(0)}_{i}(\mu) +\sum_{l\geq 1} \left(\frac{\al_s}{4\pi} \right)^{\! l \,\,}  \sum_{k=0}^{2} \al_k^L(\mu) \, V'^{(l)}_{ik}(\mu) \sqrt{z_c}  \label{Tprimeconv}\, ,
\end{align}
with $\al_0^L(\mu)\equiv 1$.
At tree-level we obtain
\begin{align}	
V^{(0)}_{1}(\mu) &= 0 \, , \quad   V'^{(0)}_{1}(\mu)  =  0 \,  ,\\
 V^{(0)}_{2}(\mu) &=1 \, , \quad
 V'^{(0)}_{2}(\mu) = 0  \,  .
\end{align}
In the following we use the abbreviations $L\equiv \log (\mu^2/m_b^2)$ and $H_{\vec{a}} (z_c) \equiv H_{\vec{a}}$ for the harmonic polylogarithms of argument $z_c$.
The one-loop results for the convoluted colour-octet kernels then read
\begin{align}
 V^{(1)}_{10}(\mu) &=   -\frac{4L}{3} + \bigg[-\frac{4 z_c^2 }{(z_c-1)^3}H_{00}+\frac{2 \left(z_c^2+10 z_c+1\right)}{3
   (z_c-1)^2}H_{1} +\frac{2  z_c (z_c+1)}{(z_c-1)^2}H_{0} \nn \\ & \quad-\frac{4  z_c
   (z_c+1)}{(z_c-1)^3}H_{2}+\pi^2\frac{2  z_c (z_c+1)}{3 (z_c-1)^3}+\frac{-5 z_c^2+18
   z_c-5}{(z_c-1)^2} \bigg] + i \pi  \bigg[ -\frac{4z_c^2}{(z_c-1)^3}H_0
   \nn \\ &\quad +\frac{2(2 z_c^2+ 5z_c-1)}{3(z_c-1)^2}\bigg] \, , \nn \\
    V^{(1)}_{11}(\mu) &=\bigg[-\frac{4 z_c^2 (z_c+3) }{(z_c-1)^4}H_{00}-\frac{2  z_c \left(z_c^2-20
   z_c-5\right)}{3 (z_c-1)^3} H_{0}-\frac{4  z_c \left(z_c^2+6
   z_c+1\right)}{(z_c-1)^4}H_{2}
   \nn \\ &\quad -\frac{2 \left(z_c^3-25 z_c^2-25 z_c+1\right)}{3
   (z_c-1)^3}  H_{1}+ \pi^2\frac{2  z_c \left(z_c^2+6 z_c+1\right)}{3 (z_c-1)^4}
    \nn \\ &\quad +\frac{-11 z_c^3+155
   z_c^2+155 z_c-11}{9 (z_c-1)^3} \bigg]
   + i \pi \bigg[-\frac{4 z_c^2 (z_c+3)}{(z_c-1)^4} H_{0}
    \nn \\ &\quad
    + \frac{-4 z_c^3+46 z_c^2+4 z_c+2}{3 (z_c-1)^3}\bigg]  \, , \nn \\
     V^{(1)}_{12}(\mu) &=\bigg[-\frac{24 z_c^2 \left(z_c^2+3 z_c+1\right) }{(z_c-1)^5}H_{00}+\frac{2  (z_c+1)^2
   \left(z_c^2+28 z_c+1\right)}{(z_c-1)^4}H_{1}
   \nn \\ &\quad +\frac{2  z_c \left(z_c^3+29 z_c^2+29
   z_c+1\right)}{(z_c-1)^4}H_{0}-\frac{24  z_c \left(z_c^3+4 z_c^2+4
   z_c+1\right)}{(z_c-1)^5}H_{2}
   \nn\\ &\quad +\pi^2\frac{4  z_c \left(z_c^3+4 z_c^2+4
   z_c+1\right)}{(z_c-1)^5}+\frac{-7 z_c^4+1368 z_c^3+4478 z_c^2+1368 z_c-7}{30 (z_c-1)^4} \bigg]
   \nn\\ &\quad + i \pi \bigg[-\frac{24  z_c^2
   \left(z_c^2+3 z_c+1\right)}{(z_c-1)^5}H_{0}+\frac{2 z_c \left(z_c^3+29 z_c^2+29 z_c+1\right)}{(z_c-1)^4}\bigg]\, ,  \nn \\
   V'^{(1)}_{10}(\mu) &= \bigg[\frac{4 z_c (z_c+2) }{3 (z_c-1)^3}H_{00}+\frac{4  \left(z_c^2+4 z_c+1\right)}{3
   (z_c-1)^3}H_{2}-\frac{2  (5 z_c+1)}{3 (z_c-1)^2}H_{0}-\frac{4
   (z_c+1)}{(z_c-1)^2}H_{1}
   \nn \\ &\quad -\pi^2\frac{2  \left(z_c^2+4 z_c+1\right)}{9 (z_c-1)^3}-\frac{4
   (z_c+1)}{(z_c-1)^2}\bigg]+ i \pi \bigg[ \frac{4 z_c (z_c+2)}{3 (z_c-1)^3} H_{0}-\frac{2 (5 z_c+1)}{3 (z_c-1)^2}\bigg] \, , \nn \\
      V'^{(1)}_{11}(\mu) &=\bigg[-\frac{4 z_c \left(z_c^2+5 z_c+2\right) }{(z_c-1)^4}H_{00}+\frac{2  \left(19
   z_c^2+28 z_c+1\right)}{3 (z_c-1)^3}H_{0}+\frac{8 \left(5 z_c^2+14 z_c+5\right)}{3
   (z_c-1)^3}H_{1}
   \nn \\ &\quad -\frac{4  \left(z_c^3+7 z_c^2+7 z_c+1\right)}{(z_c-1)^4}H_{2}+\pi^2\frac{2
   \left(z_c^3+7 z_c^2+7 z_c+1\right)}{3 (z_c-1)^4}
    \nn \\ &\quad+\frac{2 \left(41 z_c^2+206 z_c+41\right)}{9
   (z_c-1)^3}\bigg]
   + i \pi \bigg[-\frac{4  z_c \left(z_c^2+5
   z_c+2\right)}{(z_c-1)^4}H_{0}
    \nn \\ &\quad+\frac{2 \left(19 z_c^2+28 z_c+1\right)}{3 (z_c-1)^3}\bigg] \, , \nn \\
     V'^{(1)}_{12}(\mu) &=\bigg[\frac{8 z_c \left(z_c^3+10 z_c^2+12 z_c+2\right)}{(z_c-1)^5} H_{00}-\frac{2
   \left(45 z_c^3+181 z_c^2+73 z_c+1\right)}{3 (z_c-1)^4} H_{0}
   \nn\\ &\quad -\frac{4  \left(23 z_c^3+127
   z_c^2+127 z_c+23\right)}{3 (z_c-1)^4}H_{1}+\frac{8 \left(z_c^4+12 z_c^3+24 z_c^2+12
   z_c+1\right)}{(z_c-1)^5} H_{2}
   \nn \\ &\quad -\pi^2\frac{4  \left(z_c^4+12 z_c^3+24 z_c^2+12 z_c+1\right)}{3
   (z_c-1)^5}-\frac{2 \left(73 z_c^3+827 z_c^2+827 z_c+73\right)}{9 (z_c-1)^4}\bigg]
   \nn \\ &\quad + i \pi \bigg[\frac{8  z_c \left(z_c^3+10 z_c^2+12 z_c+2\right)}{(z_c-1)^5}H_{0}-\frac{2 \left(45
   z_c^3+181 z_c^2+73 z_c+1\right)}{3 (z_c-1)^4}\bigg] \, .
\end{align}
The one-loop colour-singlet kernels vanish as the corresponding colour factors are zero,
\begin{align}
   V^{(1)}_{2k}(\mu) =  V'^{(1)}_{2k}(\mu) = 0 \,  , \quad \text{for}\,\quad k=0,1,2 \, .
\end{align}
At two loops the result is rather lengthy. Here, we only present the full result for the $\mu$-dependent part which governs the scale dependence. For the $\mu$-independent part we provide a fitted function in $z_c$ that agrees with the original result at the per mill level in the range of physical values $0.05 \leq z_c \leq 0.2$. The full result is attached in electronic form to the arXiv submission of the present work. For the convoluted colour-octet kernels we obtain
\begin{align}
  V^{(2)}_{10}(\mu) &=-\frac{58}{9}L^2  + \bigg[-\frac{116 z_c^2 }{3 (z_c-1)^3}H_{00}+\frac{58  \left(z_c^2+10 z_c+1\right)}{9
   (z_c-1)^2}H_{1}+\frac{58 z_c (z_c+1)}{3 (z_c-1)^2}H_{0}
   \nn \\ & \quad-\frac{116  z_c
   (z_c+1)}{3 (z_c-1)^3}H_{2}+\pi^2\frac{58  z_c (z_c+1)}{9 (z_c-1)^3}-\frac{2 \left(527
   z_c^2-2098 z_c+527\right)}{27 (z_c-1)^2} \bigg]L
   \nn \\ &\quad + i \pi \bigg[-\frac{116  z_c^2}{3 (z_c-1)^3}H_{0}+ \frac{58 \left(2 z_c^2+5 z_c-1\right)}{9 (z_c-1)^2} \bigg]L +\bigg[  66.8297 z_c^2-43.9087 z_c
   \nn \\&\quad  -75.8620\,-\frac{0.148459}{z_c}-9.68071 \log (z_c)\bigg]+ i \pi \bigg[ -14.7418 z_c^2+37.9194 z_c
   \nn \\&\quad -23.9326\, +\frac{0.0130025}{z_c}+0.263367 \log (z_c) \bigg] \, ,\nn \\
  V^{(2)}_{11}(\mu) &= \bigg[-\frac{476 z_c^2 (z_c+3) }{9 (z_c-1)^4}H_{00}-\frac{238  z_c \left(z_c^2-20
   z_c-5\right)}{27 (z_c-1)^3}H_{0}-\frac{476 z_c \left(z_c^2+6 z_c+1\right)}{9
   (z_c-1)^4} H_{2}
   \nn \\ &\quad -\frac{238  \left(z_c^3-25 z_c^2-25 z_c+1\right)}{27 (z_c-1)^3}H_{1}+\pi^2\frac{238
    z_c \left(z_c^2+6 z_c+1\right)}{27 (z_c-1)^4}
    \nn \\ & \quad -\frac{119 \left(11 z_c^3-155
   z_c^2-155 z_c+11\right)}{81 (z_c-1)^3} \bigg]L + i \pi \bigg[ -\frac{476  (z_c+3) z_c^2}{9 (z_c-1)^4}H_{0}
   \nn \\&\quad -\frac{238 \left(2 z_c^3-23 z_c^2-2
   z_c-1\right)}{27 (z_c-1)^3} \bigg]L +\bigg[ -86.0751 z_c^2+71.1970 z_c +228.207\,
   \nn \\&\quad +\frac{0.273411}{z_c}+20.6412 \log (z_c)\bigg]+ i \pi \bigg[  -105.333 z_c -18.1118\,-\frac{0.0625952}{z_c}
   \nn \\&\quad -3.09381 \log (z_c)\bigg] \, ,\nn \\
  V^{(2)}_{12}(\mu) &= \bigg[-\frac{1096 z_c^2 \left(z_c^2+3 z_c+1\right) }{3 (z_c-1)^5}H_{00}+\frac{274
   (z_c+1)^2 \left(z_c^2+28 z_c+1\right)}{9 (z_c-1)^4}H_{1}
   \nn \\ &\quad +\frac{274  z_c \left(z_c^3+29
   z_c^2+29 z_c+1\right)}{9 (z_c-1)^4}H_{0}-\frac{1096  z_c \left(z_c^3+4 z_c^2+4
   z_c+1\right)}{3 (z_c-1)^5}H_{2}
   \nn \\ &\quad +\pi^2\frac{548  z_c \left(z_c^3+4 z_c^2+4 z_c+1\right)}{9
   (z_c-1)^5}
   \nn \\&\quad -\frac{137 \left(7 z_c^4-1368 z_c^3-4478 z_c^2-1368 z_c+7\right)}{270 (z_c-1)^4}  \bigg]L
   \nn \\ &\quad + i \pi \bigg[-\frac{1096  z_c^2
   \left(z_c^2+3 z_c+1\right)}{3 (z_c-1)^5}H_{0}+\frac{274 z_c \left(z_c^3+29 z_c^2+29 z_c+1\right)}{9 (z_c-1)^4} \bigg]L
   \nn \\ &\quad +\bigg[  -125.04 z_c^2+86.9295 z_c -26.8151\, +\frac{0.11072}{z_c}+0.595046 \log (z_c) \bigg]
   \nn \\ &\quad + i \pi \bigg[- 20.971087 z_c^3  +
  49.652981 z_c^2 - 65.251113 z_c + 32.324740  - \frac{0.054806548}{z_c}
  \nn \\&\quad + \frac{0.000082559030}{z_c^2}   + 15.519430 \log (z_c)+
  1.9371679 \log^2(z_c)\bigg] \, ,\nn \\
  V'^{(2)}_{10}(\mu) &= \bigg[ \frac{116 z_c (z_c+2) }{9 (z_c-1)^3}H_{00}+\frac{116  \left(z_c^2+4 z_c+1\right)}{9
   (z_c-1)^3}H_{2}-\frac{58 (5 z_c+1)}{9 (z_c-1)^2} H_{0}-\frac{116 (z_c+1)}{3
   (z_c-1)^2}H_{1}
   \nn \\ &\quad -\pi^2 \frac{58 \left(z_c^2+4 z_c+1\right)}{27 (z_c-1)^3}-\frac{116 (z_c+1)}{3
   (z_c-1)^2}\bigg]L + i \pi \bigg[ \frac{116  z_c (z_c+2)}{9 (z_c-1)^3}H_{0}
   \nn \\ &\quad -\frac{58 (5 z_c+1)}{9 (z_c-1)^2}\bigg]L +\bigg[ 32.231179 z_c^3   -
  58.177605 z_c^2 + 81.440153 z_c-76.877229\,
  \nn \\&\quad - \frac{0.16778620}{z_c} + \frac{0.00063267263}{z_c^2}  - 33.071605 \log (z_c) +
  0.97638808 \log^2(z_c) \bigg]
  \nn \\ &\quad + i \pi \bigg[ -30.6744 z_c^2+35.2510 z_c  -17.1594\,+\frac{0.0920248}{z_c}+1.56243 \log (z_c) \bigg] \, ,\nn \\
  V'^{(2)}_{11}(\mu) &= \bigg[-\frac{476 z_c \left(z_c^2+5 z_c+2\right) }{9 (z_c-1)^4}H_{00}+\frac{238  \left(19
   z_c^2+28 z_c+1\right)}{27 (z_c-1)^3}H_{0}
   \nn \\ &\quad +\frac{952  \left(5 z_c^2+14 z_c+5\right)}{27
   (z_c-1)^3}H_{1}-\frac{476  \left(z_c^3+7 z_c^2+7 z_c+1\right)}{9 (z_c-1)^4}H_{2}
   \nn \\ &\quad +\pi^2 \frac{238
   \left(z_c^3+7 z_c^2+7 z_c+1\right)}{27 (z_c-1)^4}+\frac{238 \left(41 z_c^2+206
   z_c+41\right)}{81 (z_c-1)^3} \bigg]L
   \nn \\ &\quad + i \pi \bigg[-\frac{476  z_c \left(z_c^2+5
   z_c+2\right)}{9 (z_c-1)^4}H_{0} +\frac{238 \left(19 z_c^2+28 z_c+1\right)}{27 (z_c-1)^3}\bigg]L
    +\bigg[ -234.47 z_c^2
    \nn \\ &\quad +316.827 z_c -148.270\,+\frac{1.51374}{z_c} -\frac{0.0135728}{z_c^2}-32.2242 \log (z_c) \bigg]
    \nn \\ &\quad + i \pi 	\bigg[  -3.48636 z_c^2-33.7104 z_c+ 37.2121\,+\frac{0.702284}{z_c} -\frac{0.00455572}{z_c^2}
    \nn \\&\quad +22.4084 \log (z_c) \bigg] \, ,\nn \\
  V'^{(2)}_{12}(\mu) &= \bigg[\frac{1096 z_c \left(z_c^3+10 z_c^2+12 z_c+2\right) }{9 (z_c-1)^5}H_{00}-\frac{274
   \left(45 z_c^3+181 z_c^2+73 z_c+1\right)}{27 (z_c-1)^4}H_{0}
   \nn \\ &\quad -\frac{548  \left(23 z_c^3+127
   z_c^2+127 z_c+23\right)}{27 (z_c-1)^4}H_{1}+\frac{1096 \left(z_c^4+12 z_c^3+24 z_c^2+12
   z_c+1\right)}{9 (z_c-1)^5}H_{2}
   \nn \\ &\quad -\pi^2\frac{548  \left(z_c^4+12 z_c^3+24 z_c^2+12
   z_c+1\right)}{27 (z_c-1)^5}-\frac{274 \left(73 z_c^3+827 z_c^2+827 z_c+73\right)}{81 (z_c-1)^4} \bigg]L
   \nn \\ &\quad + i \pi \bigg[\frac{1096  z_c \left(z_c^3+10 z_c^2+12 z_c+2\right)}{9 (z_c-1)^5}H_{0}-\frac{274 \left(45
   z_c^3+181 z_c^2+73 z_c+1\right)}{27 (z_c-1)^4} \bigg]L
   \nn \\ &\quad +\bigg[  -213.310 z_c^2+115.262 z_c +3.64826\,+\frac{4.51494}{z_c}-\frac{0.0350821}{z_c^2}
   \nn \\ &\quad +37.7768 \log (z_c) \bigg]+ i \pi \bigg[ - 120.66419 z_c^3 + 202.98069 z_c^2 - 192.44717 z_c
   \nn \\&\quad + 84.350652\,  + \frac{0.59574628}{z_c}
 - \frac{0.0025637896}{z_c^2}+ 34.327758 \log (z_c) \bigg] \, . \label{eq:2loopoctet}
\end{align}
The result for the convoluted colour-singlet kernels takes the form
\begin{align}
  V^{(2)}_{20}(\mu) &=4 L^2+ \bigg[ \frac{24 z_c^2 }{(z_c-1)^3}H_{00}-\frac{4  \left(z_c^2+10
   z_c+1\right)}{(z_c-1)^2}H_{1}-\frac{12  z_c (z_c+1)}{(z_c-1)^2}H_{0}
   \nn \\ &\quad +\frac{24
   z_c (z_c+1)}{(z_c-1)^3}H_2-\pi^2\frac{4  z_c (z_c+1)}{(z_c-1)^3}+\frac{8 \left(13
   z_c^2-44 z_c+13\right)}{3 (z_c-1)^2}\bigg]L+ i \pi \bigg[ \frac{24  z_c^2}{(z_c-1)^3}H_{0}
   \nn \\ &\quad -\frac{4 \left(2 z_c^2+5 z_c-1\right)}{(z_c-1)^2}\bigg]L +\bigg[   55.3728 z_c+ 92.1737\,+\frac{0.107621}{z_c}+5.69272 \log (z_c)\bigg]
   \nn \\ &\quad + i \pi \bigg[  -8.26434 z_c +23.3800\,-\frac{0.0109724}{z_c}-0.0317131 \log (z_c) \bigg] \, ,\nn \\
  V^{(2)}_{21}(\mu) &=  \bigg[\frac{24 z_c^2 (z_c+3) }{(z_c-1)^4}H_{00}+\frac{4  z_c \left(z_c^2-20
   z_c-5\right)}{(z_c-1)^3}H_{0}+\frac{24  z_c \left(z_c^2+6
   z_c+1\right)}{(z_c-1)^4}H_2
   \nn \\ &\quad +\frac{4  \left(z_c^3-25 z_c^2-25
   z_c+1\right)}{(z_c-1)^3}H_{1}- \pi^2\frac{4 z_c \left(z_c^2+6 z_c+1\right)}{(z_c-1)^4}
   \nn \\ &\quad +\frac{2
   \left(11 z_c^3-155 z_c^2-155 z_c+11\right)}{3 (z_c-1)^3} \bigg]L+ i \pi \bigg[ \frac{24  (z_c+3) z_c^2}{(z_c-1)^4}H_{0}
   \nn \\ &\quad +\frac{8 z_c^3-92 z_c^2-8 z_c-4}{(z_c-1)^3}\bigg]L + \bigg[  113.426 z_c -94.6182\,-\frac{0.188354}{z_c}
   \nn \\ &\quad -12.3877 \log (z_c) \bigg]+ i \pi \bigg[  -53.5714 z_c^2+48.8676 z_c+ 16.8809\,+\frac{0.0506772}{z_c}
   \nn \\ &\quad +2.03180 \log (z_c)\bigg] \, ,\nn \\
  V^{(2)}_{22}(\mu) &=  \bigg[\frac{144 z_c^2 \left(z_c^2+3 z_c+1\right) }{(z_c-1)^5}H_{00}-\frac{12  (z_c+1)^2
   \left(z_c^2+28 z_c+1\right)}{(z_c-1)^4}H_{1}
   \nn \\ &\quad -\frac{12  z_c \left(z_c^3+29 z_c^2+29
   z_c+1\right)}{(z_c-1)^4}H_{0}+\frac{144  z_c \left(z_c^3+4 z_c^2+4
   z_c+1\right)}{(z_c-1)^5}H_2
   \nn \\ &\quad - \pi^2\frac{24 z_c \left(z_c^3+4 z_c^2+4
   z_c+1\right)}{(z_c-1)^5}+\frac{7 z_c^4-1368 z_c^3-4478 z_c^2-1368 z_c+7}{5 (z_c-1)^4} \bigg]L
   \nn \\ &\quad + i \pi \bigg[ \frac{144  z_c^2 \left(z_c^2+3 z_c+1\right)}{(z_c-1)^5}H_{0}-\frac{12 z_c \left(z_c^3+29
   z_c^2+29 z_c+1\right)}{(z_c-1)^4} \bigg]L
   \nn \\ &\quad +\bigg[ -170.55583 z_c^3+246.55778 z_c^2-183.90296 z_c +23.796978\,+\frac{0.024080798}{z_c}
   \nn
   \\ &\quad+\frac{0.00019457018}{z_c^2}  +4.9343948
   \log (z_c) \bigg]+ i \pi \bigg[  23.5064 z_c -21.2951\, -\frac{0.219828}{z_c} \nn
   \\ &\quad +\frac{0.00168271}{z_c^2}-6.22465 \log (z_c)\bigg] \, ,\nn \\
  V'^{(2)}_{20}(\mu) &= \bigg[-\frac{8 z_c (z_c+2)}{(z_c-1)^3}H_{00}-\frac{8 \left(z_c^2+4
   z_c+1\right)}{(z_c-1)^3}H_{2}+\frac{4 (5 z_c+1)}{(z_c-1)^2}H_{0}+\frac{24
   (z_c+1)}{(z_c-1)^2}H_{1}
   \nn \\ &\quad +\pi^2\frac{4  \left(z_c^2+4 z_c+1\right)}{3 (z_c-1)^3}+\frac{24
   (z_c+1)}{(z_c-1)^2} \bigg]L + i \pi \bigg[\frac{4 (5 z_c+1)}{(z_c-1)^2}-\frac{8 z_c (z_c+2)}{(z_c-1)^3}H_{0} \bigg]L
   \nn \\ &\quad + \bigg[ 11.757344 z_c^2 +2.3337593 z_c -14.515538+ \frac{0.55463304}{z_c} -\frac{0.0021489969}{z_c^2}
   \nn \\ &\quad -19.152880 \log(z_c)-6.9182541 \log^2(z_c)\bigg]+ i \pi \bigg[   9.07488 z_c -8.30981\,
   \nn \\ &\quad-\frac{0.203702}{z_c}  +\frac{0.00146545}{z_c^2}-5.78268 \log (z_c)\bigg] \, ,\nn \\
  V'^{(2)}_{21}(\mu) &=  \bigg[\frac{24 z_c \left(z_c^2+5 z_c+2\right)}{(z_c-1)^4}H_{00}-\frac{4 \left(19
   z_c^2+28 z_c+1\right)}{(z_c-1)^3}H_{0}-\frac{16 \left(5 z_c^2+14
   z_c+5\right)}{(z_c-1)^3}H_{1}
   \nn\\ &\quad +\frac{24 \left(z_c^3+7 z_c^2+7
   z_c+1\right)}{(z_c-1)^4}H_{2}-\pi^2\frac{4  \left(z_c^3+7 z_c^2+7
   z_c+1\right)}{(z_c-1)^4}
   \nn \\ &\quad -\frac{4 \left(41 z_c^2+206 z_c+41\right)}{3 (z_c-1)^3} \bigg]L+ i \pi \bigg[ \frac{24 z_c \left(z_c^2+5 z_c+2\right)}{(z_c-1)^4}H_{0}
   \nn \\ &\quad -\frac{4 \left(19 z_c^2+28
   z_c+1\right)}{(z_c-1)^3}\bigg]L + \bigg[  -547.207 z_c^3+426.128 z_c^2-113.484 z_c +0.744742\,
   \nn \\ &\quad -\frac{0.212278}{z_c} \bigg]+ i \pi \bigg[-80.143647 z_c^4 +141.27932 z_c^3  -118.74488 z_c^2 +64.799529 z_c
   \nn \\ &\quad -16.163881 -\frac{0.045527818}{z_c} -4.6034803 \log (z_c)\bigg] \, ,\nn \\
  V'^{(2)}_{22}(\mu) &=  \bigg[ -\frac{48 z_c \left(z_c^3+10 z_c^2+12 z_c+2\right)}{(z_c-1)^5}H_{00}+\frac{4
   \left(45 z_c^3+181 z_c^2+73 z_c+1\right)}{(z_c-1)^4}H_{0}
   \nn \\ &\quad +\frac{8 \left(23 z_c^3+127
   z_c^2+127 z_c+23\right)}{(z_c-1)^4}H_{1}-\frac{48 \left(z_c^4+12 z_c^3+24 z_c^2+12
   z_c+1\right)}{(z_c-1)^5}H_{2}
   \nn \\ &\quad +\pi^2 \frac{8 \left(z_c^4+12 z_c^3+24 z_c^2+12
   z_c+1\right)}{(z_c-1)^5}+\frac{4 \left(73 z_c^3+827 z_c^2+827 z_c+73\right)}{3 (z_c-1)^4}\bigg]L
   \nn \\ &\quad + i \pi \bigg[\frac{4 \left(45 z_c^3+181 z_c^2+73 z_c+1\right)}{(z_c-1)^4}-\frac{48 z_c
   \left(z_c^3+10 z_c^2+12 z_c+2\right)}{(z_c-1)^5}H_{0} \bigg]L \nn \\ & \quad +\bigg[  -122.195 z_c^2+118.502 z_c -37.1904\,-\frac{0.0663899}{z_c} -\frac{0.00227343}{z_c^2}
   \nn \\ &\quad -9.22903 \log (z_c) \bigg]+ i \pi \bigg[ -170.35896 z_c^4 +201.55736 z_c^3 -93.946885 z_c^2
   \nn \\ &\quad +7.9715007 z_c +7.3620315 +\frac{0.34672826}{z_c} -\frac{0.0045037110}{z_c^2} +\frac{0.000031922024}{z_c^3}
   \nn \\ &\quad+ 4.4579847\log(z_c)\bigg] \, .
\end{align}

Finally, we checked with the full result (without interpolation in $z_c$) that in the limit $m_c\rightarrow 0$ eq.~\eqref{Tconv} with $\mu=m_b$ coincides with the result for the vertex corrections to the colour-allowed tree topology of the decay $B\to \pi\pi$ given in eq.~(48) of~\cite{Beneke:2009ek} .

\section{Conversion from the pole to the $\msbar$ scheme}
\label{sec:msbar}

The convoluted kernels in eqs.~\eqref{Tconv} and \eqref{Tprimeconv} are given in the pole scheme, where $m_c$ and $m_b$ appearing in $L\equiv \log (\mu^2/m_b^2)$ and $z_c={m_c^2}/{m_b^2}$ denote the pole quark masses, and the renormalization scale $\mu\sim m_b$. In order to discuss the scheme dependence of the convoluted kernels, we also give the results in the $\msbar$ scheme for the quark masses. Since the LO kernels are constant and the NLO colour-singlet kernels vanish, the conversion from the pole to the $\msbar$ scheme will only affect the NNLO colour-octet kernels $V_{1k}^{(2)}$ and $V_{1k}^{\prime(2)}$. To this end, using the one-loop relation between pole- and $\msbar$-quark mass,
\begin{align}
m_{q}&=\overline{m}_{q}(\mu)\left[1+\frac{\al_s}{\pi}\left(\frac{4}{3}+\log\left(\frac{\mu^2}{\overline{m}^2_q(\mu)}\right)\right)\right] \, ,
\end{align}
we find that the corresponding convoluted kernels in the $\msbar$ scheme are obtained via the relation
\begin{align} \label{eq:pole2msbar}
  V_{1k}^{(\prime)\msbar(2)}(\mu) & = V_{1k}^{(\prime)(2)} + \Delta V_{1k}^{(\prime)} \, , \nn \\
  \Delta V_{1k}& = -8z_c\ln(z_c) \frac{\partial V_{1k}^{(1)}}{\partial z_c} -\left[\frac{32}{3}+8L \right] \frac{\partial V_{1k}^{(1)}}{\partial L}\, , \nn \\
  \Delta V_{1k}^{\prime}& =-8\sqrt{z_c}\ln(z_c) \frac{\partial \sqrt{z_c}V_{1k}^{\prime(1)}}{\partial z_c} -\left[\frac{32}{3}+8L \right] \frac{\partial V_{1k}^{\prime(1)}}{\partial L}\, ,
\end{align}
where now $L\equiv\log (\mu^2/\overline{m}^2_b(\mu))$ and $z_c=\overline{m}^2_c(\mu)/\overline{m}^2_b(\mu)$, with $\mu\sim \overline{m}_b(\overline{m}_b)$.

The tree-level and one-loop kernels will have the same functional dependence as in the pole scheme, but now depend on the above new abbreviations in the $\msbar$ scheme. At two loops we explicitly give the terms that have to be added,
\begin{align}
 \Delta V_{10} &= \frac{32}{3} L + \bigg[ -\frac{32 z_c \left(z_c^2+4 z_c+1\right) }{(z_c-1)^4}H_{20}-\frac{96 z_c^2 (z_c+2)
   }{(z_c-1)^4}H_{000}+\frac{32 z_c (5 z_c+1) }{(z_c-1)^3}H_{00} \nn \\ &\quad
   +\frac{96 z_c
   (z_c+1) }{(z_c-1)^3}H_{10}-\frac{32  z_c \left(z_c^2-11 z_c-8\right)}{3
   (z_c-1)^3}H_{0}-\frac{64  z_c \left(z_c^2+4 z_c+1\right)}{(z_c-1)^4}H_{3} \nn \\ &\quad
   +\frac{96
   z_c (z_c+1)}{(z_c-1)^3}H_{2}+\pi^2\frac{16   z_c \left(z_c^2+4
   z_c+1\right)}{3 (z_c-1)^4}H_{0} +\frac{128}{9}\bigg]+ i \pi \bigg[ \frac{16  z_c (5 z_c+1)}{(z_c-1)^3}H_{0} \nn \\ &\quad
   -\frac{64 z_c^2 (z_c+2) }{(z_c-1)^4}H_{00}\bigg] \, , \nn \\
 \Delta V_{11} &=  \bigg[\frac{32 z_c \left(23 z_c^2+68 z_c+5\right) }{3 (z_c-1)^4}H_{00}+\frac{64 z_c \left(7
   z_c^2+34 z_c+7\right) }{3 (z_c-1)^4}H_{10} \nn \\ &\quad
   -\frac{96 z_c^2 \left(z_c^2+9 z_c+6\right)
   }{(z_c-1)^5}H_{000}-\frac{32 z_c \left(z_c^3+15 z_c^2+15 z_c+1\right)
   }{(z_c-1)^5}H_{20}  \nn \\ &\quad
   +\frac{16  z_c \left(73 z_c^2+430 z_c+73\right)}{9
   (z_c-1)^4}H_{0}+\frac{64  z_c \left(7 z_c^2+34 z_c+7\right)}{3 (z_c-1)^4}H_{2}  \nn \\ &\quad
   -\frac{64
   z_c \left(z_c^3+15 z_c^2+15 z_c+1\right)}{(z_c-1)^5}H_{3} + \pi^2\frac{16  z_c \left(z_c^3+15 z_c^2+15
   z_c+1\right)}{3 (z_c-1)^5}H_{0} \bigg]  \nn \\ &\quad
   +i \pi \bigg[\frac{16 H_{0} z_c \left(23 z_c^2+68 z_c+5\right)}{3 (z_c-1)^4}-\frac{64 z_c^2 \left(z_c^2+9
   z_c+6\right) H_{00}}{(z_c-1)^5}\bigg] \, , \nn \\
 \Delta V_{12} &=\bigg[\frac{32 z_c \left(45 z_c^3+181 z_c^2+73 z_c+1\right) }{(z_c-1)^5}H_{00}+\frac{32 z_c
   \left(23 z_c^3+127 z_c^2+127 z_c+23\right) }{(z_c-1)^5}H_{10} \nn \\ &\quad
   -\frac{576 z_c^2
   \left(z_c^3+10 z_c^2+12 z_c+2\right) }{(z_c-1)^6}H_{000}+\frac{32  z_c \left(23 z_c^3+127
   z_c^2+127 z_c+23\right)}{(z_c-1)^5}H_{2}\nn \\ &\quad-\frac{192 z_c \left(z_c^4+12
   z_c^3+24 z_c^2+12 z_c+1\right) }{(z_c-1)^6}H_{20} \nn \\ &\quad
   -\frac{384  z_c \left(z_c^4+12 z_c^3+24
   z_c^2+12 z_c+1\right)}{(z_c-1)^6}H_{3} \nn \\&\quad
   +\frac{16  z_c \left(73
   z_c^3+827 z_c^2+827 z_c+73\right)}{3 (z_c-1)^5}H_{0} \nn \\ &\quad
   +\pi^2\frac{32   z_c
   \left(z_c^4+12 z_c^3+24 z_c^2+12 z_c+1\right)}{(z_c-1)^6}H_{0}\bigg] \nn \\&\quad
   + i\pi \bigg[ \frac{16  z_c \left(45 z_c^3+181 z_c^2+73 z_c+1\right)}{(z_c-1)^5}H_{0}-\frac{384 z_c^2
   \left(z_c^3+10 z_c^2+12 z_c+2\right) }{(z_c-1)^6}H_{00}\bigg] \, ,
\end{align}
\begin{align}
   \Delta V^{\prime}_{10} &= \bigg[-\frac{16 \left(9 z_c^2+26 z_c+1\right) }{3 (z_c-1)^3}H_{00}-\frac{16 \left(5 z_c^2+26
   z_c+5\right) }{3 (z_c-1)^3}H_{10} -\frac{64 H_{0} \left(z_c^2+7 z_c+1\right)}{3 (z_c-1)^3}\nn \\ &\quad
   +\frac{16 z_c \left(z_c^2+11 z_c+6\right)
   }{(z_c-1)^4}H_{000}+\frac{16 \left(z_c^3+17 z_c^2+17 z_c+1\right) }{3
   (z_c-1)^4}H_{20} \nn \\ &\quad
  -\frac{16  \left(5
   z_c^2+26 z_c+5\right)}{3 (z_c-1)^3}H_{2}+\frac{32  \left(z_c^3+17 z_c^2+17
   z_c+1\right)}{3 (z_c-1)^4}H_{3} \nn \\ &\quad   - \pi^2\frac{8   \left(z_c^3+17 z_c^2+17 z_c+1\right)}{9
   (z_c-1)^4}H_{0}\bigg] + i \pi \bigg[ \frac{32 z_c \left(z_c^2+11 z_c+6\right) }{3 (z_c-1)^4}H_{00} \nn \\ &\quad
   -\frac{8  \left(9
   z_c^2+26 z_c+1\right)}{3 (z_c-1)^3}H_{0}\bigg] \nn \, , \\
   \Delta V^{\prime}_{11} &= \bigg[ \frac{16 \left(31 z_c^3+239 z_c^2+113 z_c+1\right) }{3 (z_c-1)^4}H_{00}+\frac{256 \left(z_c^3+11
   z_c^2+11 z_c+1\right) }{3 (z_c-1)^4}H_{10} \nn \\ &\quad
   -\frac{48 z_c \left(z_c^3+22 z_c^2+35
   z_c+6\right) }{(z_c-1)^5}H_{000}-\frac{16 \left(z_c^4+28 z_c^3+70 z_c^2+28 z_c+1\right)
   }{(z_c-1)^5}H_{20}  \nn \\ &\quad
   +\frac{8  \left(47 z_c^3+1105 z_c^2+1105 z_c+47\right)}{9
   (z_c-1)^4}H_{0}+\frac{256  \left(z_c^3+11 z_c^2+11 z_c+1\right)}{3 (z_c-1)^4}H_{2} \nn \\ &\quad -\frac{32
    \left(z_c^4+28 z_c^3+70 z_c^2+28 z_c+1\right)}{(z_c-1)^5}H_{3} \nn \\ &\quad
    +\pi^2 \frac{8  \left(z_c^4+28 z_c^3+70 z_c^2+28
   z_c+1\right)}{3 (z_c-1)^5}H_{0}\bigg]
   \nn \\ &\quad + i \pi \bigg[ \frac{8  \left(31 z_c^3+239 z_c^2+113 z_c+1\right)}{3 (z_c-1)^4}H_{0}-\frac{32 z_c
   \left(z_c^3+22 z_c^2+35 z_c+6\right) }{(z_c-1)^5}H_{00}\bigg] \nn \, , \\
   \Delta V^{\prime}_{12} &= \bigg[-\frac{16 \left(69 z_c^4+1098 z_c^3+1558 z_c^2+274 z_c+1\right) }{3 (z_c-1)^5}H_{00} \nn \\ &\quad
      -\frac{16
   \left(35 z_c^4+686 z_c^3+1558 z_c^2+686 z_c+35\right) }{3 (z_c-1)^5}H_{10} \nn \\ &\quad
   +\frac{96 z_c
   \left(z_c^4+39 z_c^3+130 z_c^2+74 z_c+6\right) }{(z_c-1)^6}H_{000} \nn \\ &\quad
   +\frac{32
   \left(z_c^5+45 z_c^4+204 z_c^3+204 z_c^2+45 z_c+1\right) }{(z_c-1)^6}H_{20} \nn \\ &\quad
   -\frac{8  \left(79 z_c^4+3700 z_c^3+10442 z_c^2+3700 z_c+79\right)}{9
   (z_c-1)^5}H_{0} \nn \\&\quad
   -\frac{16  \left(35 z_c^4+686 z_c^3+1558 z_c^2+686 z_c+35\right)}{3
   (z_c-1)^5}H_{2}
    \nn \\ &\quad +\frac{64  \left(z_c^5+45 z_c^4+204 z_c^3+204 z_c^2+45
   z_c+1\right)}{(z_c-1)^6}H_{3}
   \nn \\ &\quad -\pi^2\frac{16
     \left(z_c^5+45 z_c^4+204 z_c^3+204 z_c^2+45 z_c+1\right)}{3
   (z_c-1)^6}H_{0} \bigg] \nn \\ &\quad
   + i \pi \bigg[\frac{64 z_c \left(z_c^4+39 z_c^3+130 z_c^2+74 z_c+6\right) }{(z_c-1)^6}H_{00}
   \nn \\ &\quad -\frac{8
    \left(69 z_c^4+1098 z_c^3+1558 z_c^2+274 z_c+1\right)}{3 (z_c-1)^5}H_{0} \bigg] \, .
\end{align}

\section{Phenomenological applications}
\label{sec:pheno}

In this section we perform an extensive phenomenological analysis of $\bar{B}_{(s)}\to D_{(s)}^{(\ast)+} \, L^-$ and $\Lambda_b \to \Lambda_c^+ \, L^-$ decays in QCDF. Like before, $L$ is a light meson from the set $\{\pi,\rho,K^{(\ast)},a_1\}$. We take into account the expressions through to NNLO for the hard scattering kernels, and the most recent values for non-perturbative input parameters, which we specify below. We analyze the impact of the NNLO correction on the topological tree amplitude $a_1(D^{(\ast)+}L^{-})$, and subsequently predict the branching ratios for the mesonic decays. Afterwards, we perform tests of QCD factorization by considering suitably chosen ratios of non-leptonic to either semi-leptonic or non-leptonic channels. Finally, we give the theoretical predictions for baryonic decays.

\subsection{Input parameters}
\label{subsec:input}

Here we collect in Table~\ref{tab:inputs} the theoretical input parameters entering our numerical analysis throughout this paper. They include the SM parameters such as the CKM matrix elements, quark masses, and the strong coupling constant, as well as the hadronic parameters such as meson decay constants, transition form factors, and the Gegenbauer moments of light mesons. Three-loop running is used for $\alpha_s$ throughout this paper. Furthermore, we use a two-loop relation between pole and $\MSbar$ mass to convert the top-quark pole mass $m_t^{\text{pole}}$ to the scale-invariant mass $\overline{m}_t(\overline{m}_t)$~\cite{Chetyrkin:2000yt}.

\begin{table}[htbp]
\begin{center}
\caption{\label{tab:inputs} Summary of theoretical input parameters. The Gegenbauer moments of light mesons are evaluated at $\mu=1~\gev$.}
\vspace*{0.0cm}
\renewcommand{\arraystretch}{1.5}
{\tabcolsep=0.638cm\begin{tabular}{|cccc|c|}
\hline\hline
\multicolumn{5}{|l|}{\textbf{QCD and electroweak parameters}}
\\
\hline
  $G_F [10^{-5}\gev^{-2}]$
& $\alpha_s(m_Z)$
& $m_Z [\gev]$
& $m_W [\gev]$
&
\\
  $1.1663787$
& $0.1185 \pm 0.0006$
& $91.1876$
& $80.385$
& \cite{Agashe:2014kda}
\\
\hline
\end{tabular}}
{\tabcolsep=0.122cm \begin{tabular}{|cccccc|c|}
\hline
\multicolumn{7}{|l|}{\hspace{0.35cm} \textbf{Quark masses [GeV]}}
\\
\hline
  $m_t^{\rm pole}$
& $m_b^{\rm pole}$
& $m_c^{\rm pole}$
& $\overline{m}_t(\overline{m}_t)$
& $\overline{m}_b(\overline{m}_b)$
& $\overline{m}_c(\overline{m}_c)$
&
\\
  $173.34 \pm 0.76$
& $4.78 \pm 0.06$
& $1.67 \pm 0.07$
& $163.99 \pm 0.72$
& $4.18  \pm 0.03$
& $1.275 \pm 0.025$
& \cite{Agashe:2014kda,ATLAS:2014wva}
\\
\hline
\end{tabular}}
{\tabcolsep=0.659cm \begin{tabular}{|ccc|c|}
\hline
\multicolumn{4}{|l|}{\hspace{-0.3cm} \textbf{CKM matrix elements}}
\\
\hline
  $|V_{ud}|$
& $|V_{us}|$
& $|V_{cb}|_{\rm exclusive} [10^{-3}]$
&
\\
  $0.97417 \pm 0.00021$
& $0.2253  \pm 0.0008 $
& $39.5 \pm 0.8$
& \cite{Agashe:2014kda,Hardy:2014qxa,Charles:2004jd}
\\
\hline
\end{tabular}}
{\tabcolsep=0.151cm \begin{tabular}{|cccccc|c|}
\hline
\multicolumn{7}{|l|}{\hspace{0.38cm} \textbf{Lifetimes and masses of $B_{d,s}$ and $\Lambda_b$}}
\\
\hline
  $\tau_{B_d} [{\rm ps}]$
& $\tau_{B_s} [{\rm ps}]$
& $\tau_{\Lambda_b} [{\rm ps}]$
& $m_{B_d} [\mev]$
& $m_{B_s} [\mev]$
& $m_{\Lambda_b} [\mev]$
&
\\
  $1.520 \pm 0.004$
& $1.505 \pm 0.004$
& $1.466 \pm 0.010$
& $5279.61$
& $5366.79$
& $5619.51$
& \cite{Agashe:2014kda,Amhis:2014hma}
\\
\hline
\end{tabular}}
{\tabcolsep=0.109cm \begin{tabular}{|cccccc|c|}
\hline
\multicolumn{7}{|l|}{\hspace{0.40cm} \textbf{$\boldsymbol{B\to D^{(\ast)}}$ transition form factors}}
\\
\hline
\multicolumn{1}{|c|}{}
& $F(1)|V_{cb}| [10^{-3}]$
& $\rho^2$
& $R_1$
& $R_2$
& $R_3$
&
\\
\hline
\multicolumn{1}{|c|}{$B\to D$}
& $42.65 \pm 1.53$
& $1.185 \pm 0.054$
& --
& --
& --
& \cite{Amhis:2014hma,Sakaki:2012ft}
\\
\hline
\multicolumn{1}{|c|}{$B\to D^{\ast}$}
& $35.81 \pm 0.45$
& $1.207 \pm 0.026$
& $1.406 \pm 0.033$
& $0.853 \pm 0.020$
& $0.97 \pm 0.10$
& \cite{Amhis:2014hma,Fajfer:2012vx}
\\
\hline
\end{tabular}}
{\tabcolsep=0.324cm \begin{tabular}{|cccccc|c|}
\hline
\multicolumn{7}{|l|}{\hspace{0.30cm} \textbf{$\boldsymbol{B_s\to D_s^{(\ast)}}$ transition form factors}}
\\
\hline
\multicolumn{1}{|c|}{}
& $F_+$
& $F_0$
& $A_0$
& $A_1$
& $A_2$
&
\\
\hline
\multicolumn{1}{|c|}{$F(0)$}
& $0.7 \pm 0.1$
& $0.7 \pm 0.1$
& $0.52 \pm 0.06$
& $0.62 \pm 0.01$
& $0.75 \pm 0.07$
& \cite{Blasi:1993fi}
\\
\hline
\multicolumn{1}{|c|}{${\rm M_{res}} [\gev]$}
& $6.3$
& $6.8$
& $6.3$
& $6.8$
& $6.8$
& \cite{Blasi:1993fi}
\\
\hline
\end{tabular}}
{\tabcolsep=0.170cm \begin{tabular}{|cccccc|c|}
\hline
\multicolumn{7}{|l|}{\hspace{0.42cm} \textbf{Light-meson decay constants and Gegenbauer moments}}
\\
\hline
\multicolumn{1}{|c|}{}
& $\pi$
& $K$
& $\rho$
& $K^{\ast}$
& $a_1(1260)$
&
\\
\hline
\multicolumn{1}{|c|}{$f_L [\mev]$}
& $130.2 \pm 1.4$
& $155.6 \pm 0.4$
& $216 \pm 6$
& $211 \pm 7$
& $238 \pm 10$
& \cite{Rosner:2015wva,Straub:2015ica,Dimou:2012un,Yang:2007zt}
\\
\hline
\multicolumn{1}{|c|}{$\alpha_1^L$}
& --
& $-0.07 \pm 0.04$
& --
& $-0.06 \pm 0.04$
& --
&
\\
\multicolumn{1}{|c|}{$\alpha_2^L$}
& $0.29 \pm 0.08$
& $0.24 \pm 0.08$
& $0.17 \pm 0.07$
& $0.16 \pm 0.09$
& $-0.02 \pm 0.02$
& \cite{Straub:2015ica,Dimou:2012un,Yang:2007zt,Arthur:2010xf}
\\
\hline\hline
\end{tabular}}
\renewcommand{\arraystretch}{1.0}
\end{center}
\end{table}

For the $B\to D^{(\ast)}$ transition form factors, we adopt the parameterization proposed by Caprini, Lellouch, and Neubert~(CLN)~\cite{Caprini:1997mu}, with the relevant parameters extracted from exclusive semileptonic $b\to c \ell \nu_{\ell}$ decays~\cite{Amhis:2014hma}. For the $B_s\to D_s^{(\ast)}$ transition form factors, on the other hand, we use the results obtained by QCD sum-rule techniques, assuming a polar dependence on $q^2$ that is dominated by the nearest resonance~\cite{Blasi:1993fi,Colangelo:1992cx}. However, to discuss the ${\rm SU(3)}$-breaking effects in the form-factor and decay-constant ratios, we adopt the most recent lattice QCD results for the ratios~\cite{Bailey:2012rr,Rosner:2015wva}
\begin{align}
\frac{F_0^{B_s\to D_s}(m_{\pi}^2)}{F_0^{B\to D}(m_{\pi}^2)}&=1.054\pm 0.047_{\rm stat.} \pm 0.017_{\rm syst.}\,, \nonumber\\
\frac{F_0^{B_s\to D_s}(m_{\pi}^2)}{F_0^{B\to D}(m_{K}^2)}&=1.046\pm 0.044_{\rm stat.} \pm 0.015_{\rm syst.}\,, \nonumber\\
\frac{f_K}{f_\pi} & = 1.1927 \pm 0.0026 \, .
\end{align}
Neither of the form-factor ratios shows significant deviation from the U-spin symmetry.

For the $\Lambda_b\to \Lambda_c$ transition form factors, we use the most recent high-precision lattice QCD calculation with $2 + 1$ dynamical flavours~\cite{Detmold:2015aaa}. Here the $q^2$ dependence of the form factors is parameterized in a simplified $z$ expansion~\cite{Bourrely:2008za}, modified to account for pion-mass and lattice-spacing dependence. All relevant formulas and input data can be found in eq.~(79) and Tables~VII~--~IX of~\cite{Detmold:2015aaa}. Following the procedure recommended in~\cite{Detmold:2015aaa}, we calculate the central value, statistical uncertainty, and total systematic uncertainty of any observable depending on the form-factor parameters according to eqs.~(82)~--~(84) in ~\cite{Detmold:2015aaa}. Furthermore, we have also taken into account the correlation matrices between the form-factor parameters.

The decay constants $f_{\pi}$ and $f_K$ are averaged over the two-flavour lattice QCD results~\cite{Rosner:2015wva}, while $f_{\rho}$ and $f_{K^{\ast}}$ are determined from experiments~\cite{Straub:2015ica}. The light-meson Gegenbauer moments are determined by the QCD sum rule approach~\cite{Straub:2015ica,Dimou:2012un} and the lattice QCD calculation~\cite{Arthur:2010xf}. For the hadronic inputs of the axial-vector meson $a_1(1260)$, we use the results presented in ref.~\cite{Yang:2007zt}. It is noted that the Gegenbauer moments are evaluated at $\mu=1~\gev$, and are evolved to the characteristic scale $\mu\sim m_b$~\cite{Mueller:1993hg,Mikhailov:1984ii,Mueller:1994cn}. We use LL running of the Gegenbauer moments for the tree-level and the one-loop amplitude, but NLL running in the two-loop amplitude. Moreover, the running of the Gegenbauer moments is performed in the four-flavour scheme.

\subsection{Predictions for $\boldsymbol{a_1(D^{(\ast)+}L^-)}$}
\label{subsec:a1DL}

\begin{figure}[t]
\centerline{
\includegraphics[width=8.5cm]{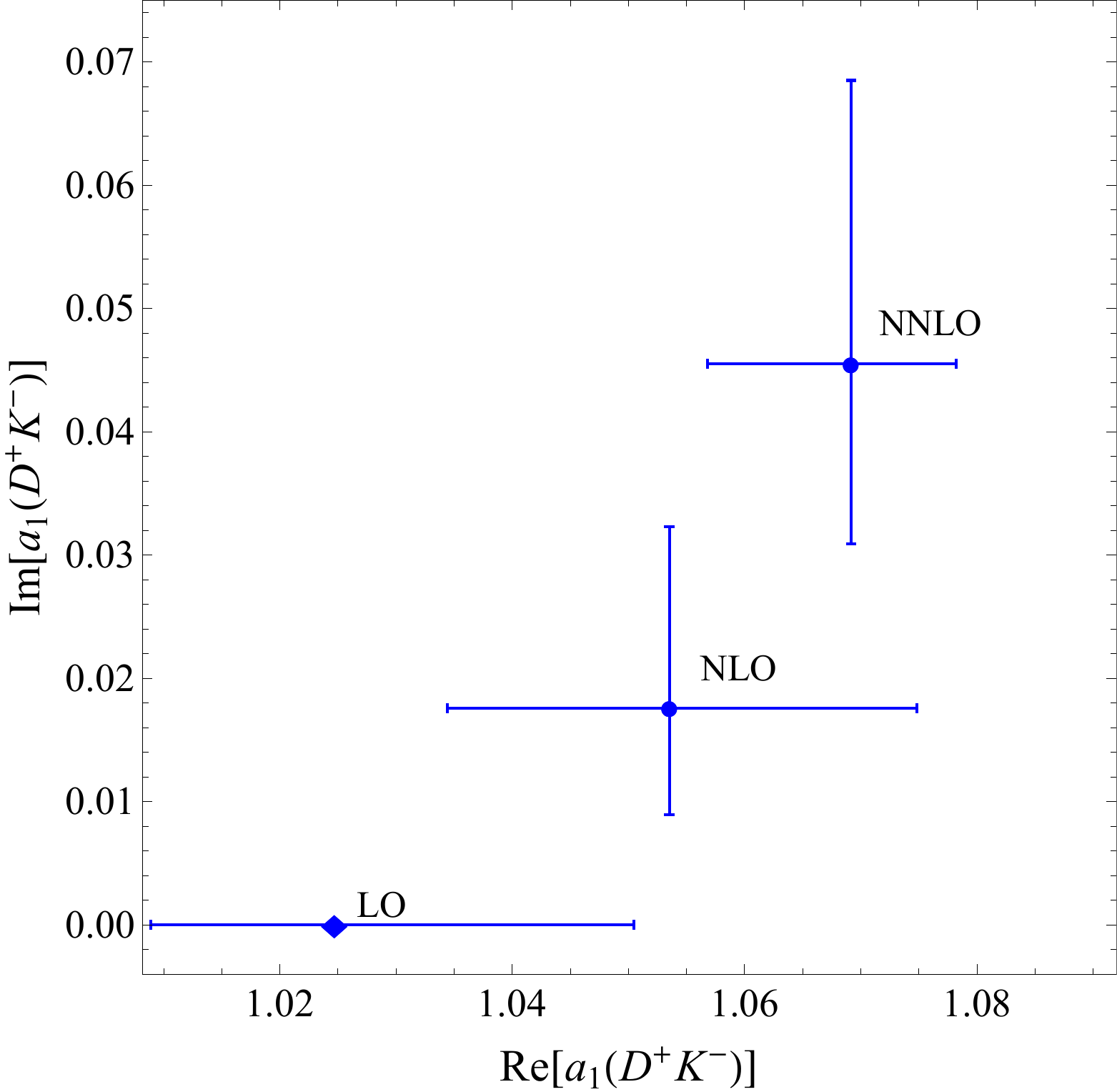}}
\caption{\label{a1DKplot} Graphical representation of $a_1(D^+ K^-)$ in the complex plane at LO, NLO and NNLO. The theoretical error estimates are also indicated.}
\end{figure}

We are now in the position to perform a numerical analysis of the coefficients $a_1(D^{(\ast)+}L^-)$ according to the expressions
\begin{align}
a_1(D^{+}L^-) & = \sum_{i=1}^2 \, C_i(\mu) \, \int_0^1 \!\! du \; \left[ \hat T_{i}(u,\mu) + \hat T'_{i}(u,\mu)\right] \, \Phi_L(u,\mu) \, , \nonumber \\
a_1(D^{*+}L^-) & = \sum_{i=1}^2 \, C_i(\mu) \, \int_0^1 \!\! du \; \left[ \hat T_{i}(u,\mu) - \hat T'_{i}(u,\mu)\right] \, \Phi_L(u,\mu) \, ,
\end{align}
into which eqs.~\eqref{Tconv} and~\eqref{Tprimeconv} have to be inserted. Using the NNLO Wilson coefficients $C_i(\mu)$ in the CMM basis~\cite{Gorbahn:2004my}, together with the input parameters collected in Table~\ref{tab:inputs}, our final numerical results for $a_1(D^+K^-)$ are given as
\begin{eqnarray}
a_1(D^+K^-) &=& 1.025 + [0.029 + 0.018 i]_{\rm NLO} + [0.016 + 0.028i]_{\rm NNLO}\, \no \\
&=& (1.069^{+0.009}_{-0.012}) + (0.046^{+0.023}_{-0.015})i\,,
\label{eq:a1DK}
\end{eqnarray}
where the number without bracket is the LO contribution, which has no imaginary part, and the following two numbers are the NLO and NNLO terms, respectively. The total errors comprise the uncertainties, added in quadrature, from the variation of the scales $\mu \in [m_b/2,2m_b]$ and $\mu_0\in [m_W/2,2m_W]$, the quark masses, the Gegenbauer moments, and $\alpha_s(m_Z)$. Unless stated otherwise, the numbers given here and below are obtained with the $b$- and $c$-quark masses renormalized in the pole scheme, which is set as our default scheme. It is observed that both the NLO and NNLO contributions add always constructively to the LO result. We also observe that the new two-loop correction is quite small in the real, but rather large in the imaginary part. It amounts to approximately $60\%$ ($2\%$) of the total imaginary (real) part of $a_1(D^+K^-)$. We emphasize that the sizable NNLO correction to the imaginary part does not indicate a breakdown of the perturbative expansion, but is due to the fact that the imaginary part vanishes at LO, and its NLO term is colour suppressed and proportional to the small Wilson coefficient $C_1(\mu)$. Moreover, the impact of the imaginary part on $|a_1(D^+K^-)|$ is only marginal. Graphical representations of $a_1(D^+K^-)$ are shown in figure~\ref{a1DKplot} at LO, NLO and NNLO.

\begin{figure}[t]
\centerline{
\hspace*{-8pt}\includegraphics[width=16.2cm]{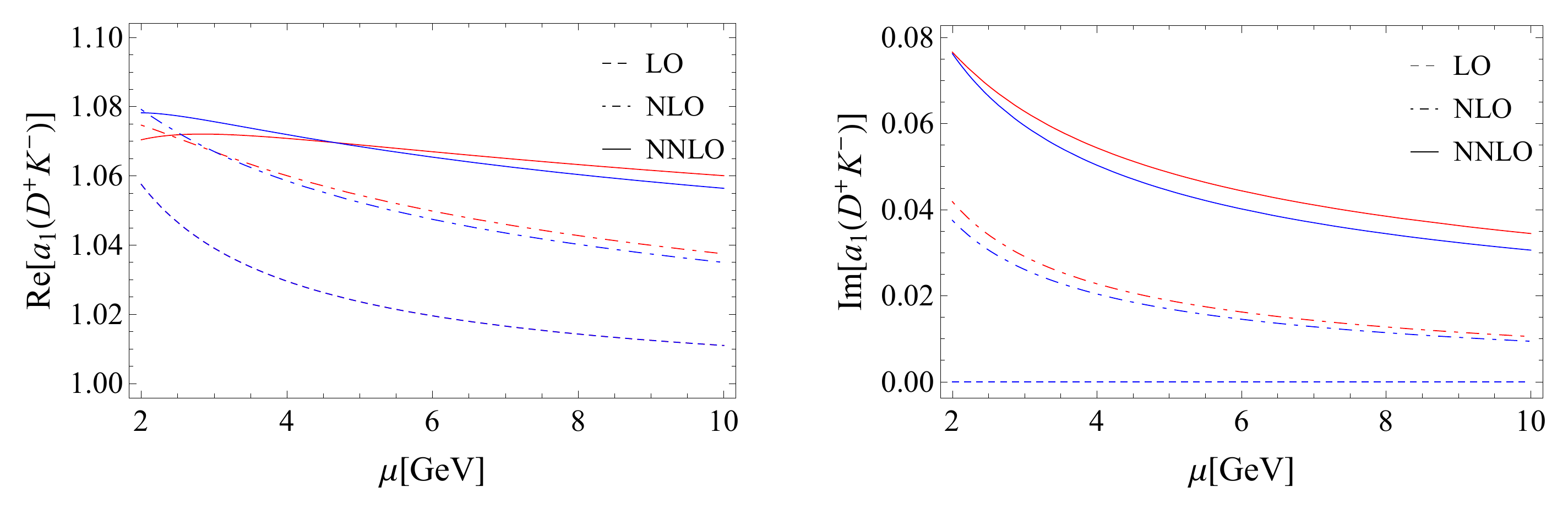}}
\caption{\label{fig:scale_dep_a1_DK} The dependence of the coefficient $a_1(D^+ K^-)$ on the renormalization scale $\mu$ both in the pole~(blue) and in the $\msbar$~(red) scheme for $b$- and $c$-quark masses. Dashed, dashed-dotted and solid lines represent the LO, NLO, and NNLO results, respectively.}
\end{figure}

Due to the truncation of the perturbative expansion, the obtained values in eq.~\eqref{eq:a1DK} depend on the renormalization scale $\mu$, which is usually considered as a measure of the accuracy of the approximation at a given order in the perturbative expansion. This is shown in figure~\ref{fig:scale_dep_a1_DK} for $a_1(D^+ K^-)$ up to NNLO, where results both in the pole~(blue) and in the $\msbar$~(red) scheme for $b$- and $c$-quark masses are given. We observe a pronounced stabilization of the scale dependence for the real part, but not for the imaginary part. This is again explained by the fact that the imaginary part vanishes at LO. It is also observed that the dependence on the $b$- and $c$-quark mass scheme is quite small, especially for the real part. We finally remark that also within a given quark-mass scheme the dependence of $a_1(D^+K^-)$ on the value of $z_c$ is minor. The dependence of $a_1(D^+K^-)$ on the second Gegenbauer moment is small, too.

It is also interesting to mention that, even up to NNLO, the coefficients $a_1(D^{(\ast)+}L^-)$ are quasi-universal, with very small process-dependent non-factorizable corrections, a fact that was observed already at NLO in ref.~\cite{Beneke:2000ry}. This is clearly seen from the following numerical results for different final states:
\begin{align} \label{eq:a1_compare}
&a_1(D^{+}K^{-}) = (1.069^{+0.009}_{-0.012}) + (0.046^{+0.023}_{-0.015})i\,,\no \\
&a_1(D^{+}\pi^{-}) = (1.072^{+0.011}_{-0.013}) + (0.043^{+0.022}_{-0.014})i\,,\no \\
&a_1(D^{\ast+}K^{-}) = (1.068^{+0.010}_{-0.012}) + (0.034^{+0.017}_{-0.011})i\,,\no \\
&a_1(D^{\ast+}\pi^{-}) = (1.071^{+0.012}_{-0.013}) + (0.032^{+0.016}_{-0.010})i\,.
\end{align}

\subsection{Predictions for class-I decays}
\label{subsec:brs}

It is generally believed that the factorization theorem is well established in class-I decays of the form $\bar{B}_{(s)}\to D_{(s)}^{(\ast)+}L^-$, where the spectator anti-quark of the initial $\bar{B}_{(s)}$ mesons is absorbed only by the $D_{(s)}^{(\ast)+}$ mesons~\cite{Beneke:2000ry,Bauer:2001cu}. We now present in Table~\ref{tab:br} our predictions for the branching ratios of these decays through to NNLO. The explicit formulas for the branching ratios can be found in~\cite{Beneke:2000ry} and shall not be repeated here. The experimental data is taken from the Particle Data Group~(PDG)~\cite{Agashe:2014kda} and/or the Heavy Flavor Averaging Group~(HFAG)~\cite{Amhis:2014hma}. For the vector and axial-vector final states, the results refer to the longitudinal polarization amplitudes only, with the longitudinal polarization fractions taken from \cite{Csorna:2003bw} for $\bar{B}_d\to D^{\ast+}\rho^-$ and \cite{Louvot:2010rd} for $\bar{B}_s\to D_s^{\ast+}\rho^-$, respectively.

\begin{table}[htbp]
\tabcolsep0.53cm
\let\oldarraystretch=\arraystretch
\renewcommand*{\arraystretch}{1.1}
\begin{center}
\caption{\label{tab:br} CP-averaged branching ratios (in units of $10^{-3}$ for $b\to c\bar{u}d$ and $10^{-4}$ for $b\to c\bar{u}s$ transitions) of $\bar{B}_{(s)}\to D_{(s)}^{(\ast)+}L^-$ decays. The vector- and axial-vector final states refer to the longitudinal polarization amplitudes only. The theoretical errors shown correspond to the uncertainties due to renormalization scales $\mu$ and $\mu_0$, the CKM as well as the hadronic parameters, added in quadrature. The experimental data is taken from refs.~\cite{Agashe:2014kda,Amhis:2014hma,Csorna:2003bw,Louvot:2010rd}.}
\vspace{0.2cm}
\begin{tabular}{lccccc}
\hline \hline
&&&&\\[-0.5cm]
Decay mode & ${\rm LO}$ & ${\rm NLO}$ & ${\rm NNLO}$ & Exp. \\
\hline
&&&&\\[-0.3cm]
  $\bar{B}_d\to D^+\pi^-$
& $\phantom{-}3.58$
& $\phantom{-}3.79_{\,-0.42}^{\,+0.44}$
& $\phantom{-}3.93_{\,-0.42}^{\,+0.43}$
& $\phantom{-}2.68\pm0.13$ \\ \addlinespace
  $\bar{B}_d\to D^{\ast+}\pi^-$
& $\phantom{-}3.15$
& $\phantom{-}3.32_{\,-0.49}^{\,+0.52}$
& $\phantom{-}3.45_{\,-0.50}^{\,+0.53}$
& $\phantom{-}2.76\pm0.13$ \\ \addlinespace
  $\bar{B}_d\to D^{+}\rho^-$
& $\phantom{-}9.51$
& $\phantom{0}10.06_{\,-1.19}^{\,+1.25}$
& $\phantom{0}10.42_{\,-1.20}^{\,+1.24}$
& $\phantom{-}7.5\pm1.2$ \\ \addlinespace
  $\bar{B}_d\to D^{\ast+}\rho^-$
& $\phantom{-}8.45$
& $\phantom{-}8.91_{\,-0.71}^{\,+0.74}$
& $\phantom{-}9.24_{\,-0.71}^{\,+0.72}$
& $\phantom{-}6.0\pm0.8$ \\ \addlinespace
\hline
&&&&\\[-0.3cm]
  $\bar{B}_s\to D_s^+\pi^-$
& $\phantom{-}4.00$
& $\phantom{-}4.24_{\,-1.15}^{\,+1.32}$
& $\phantom{-}4.39_{\,-1.19}^{\,+1.36}$
& $\phantom{-}3.04\pm0.23$ \\ \addlinespace
  $\bar{B}_s\to D_s^{\ast+}\pi^-$
& $\phantom{-}2.05$
& $\phantom{-}2.16_{\,-0.49}^{\,+0.54}$
& $\phantom{-}2.24_{\,-0.50}^{\,+0.56}$
& $\phantom{-}2.0\pm0.5$ \\ \addlinespace
  $\bar{B}_s\to D_s^{+}\rho^-$
& $\phantom{0}10.31$
& $\phantom{0}10.91_{\,-3.02}^{\,+3.46}$
& $\phantom{0}11.30_{\,-3.11}^{\,+3.56}$
& $\phantom{-}7.0\pm1.5$ \\ \addlinespace
  $\bar{B}_s\to D_s^{\ast+}\rho^-$
& $\phantom{-}5.86$
& $\phantom{-}6.18_{\,-1.28}^{\,+1.38}$
& $\phantom{-}6.41_{\,-1.31}^{\,+1.42}$
& $\phantom{0}10.2\pm2.5$ \\ \addlinespace
\hline
&&&&\\[-0.3cm]
  $\bar{B}_d\to D^+K^-$
& $\phantom{-}2.74$
& $\phantom{-}2.90_{\,-0.31}^{\,+0.33}$
& $\phantom{-}3.01_{\,-0.31}^{\,+0.32}$
& $\phantom{-}1.97\pm0.21$ \\ \addlinespace
  $\bar{B}_d\to D^{\ast+}K^-$
& $\phantom{-}2.37$
& $\phantom{-}2.50_{\,-0.36}^{\,+0.39}$
& $\phantom{-}2.59_{\,-0.37}^{\,+0.39}$
& $\phantom{-}2.14\pm0.16$ \\ \addlinespace
  $\bar{B}_d\to D^{+}K^{\ast-}$
& $\phantom{-}4.79$
& $\phantom{-}5.07_{\,-0.62}^{\,+0.65}$
& $\phantom{-}5.25_{\,-0.63}^{\,+0.65}$
& $\phantom{-}4.5\pm0.7$ \\ \addlinespace
  $\bar{B}_d\to D^{\ast+}K^{\ast-}$
& $\phantom{-}4.30$
& $\phantom{-}4.54_{\,-0.40}^{\,+0.41}$
& $\phantom{-}4.70_{\,-0.39}^{\,+0.40}$
& -- \\ \addlinespace
\hline
&&&&\\[-0.3cm]
  $\bar{B}_s\to D_s^+K^-$
& $\phantom{-}3.05$
& $\phantom{-}3.23_{\,-0.88}^{\,+1.01}$
& $\phantom{-}3.34_{\,-0.90}^{\,+1.04}$
& -- \\ \addlinespace
  $\bar{B}_s\to D_s^{\ast+}K^-$
& $\phantom{-}1.53$
& $\phantom{-}1.61_{\,-0.36}^{\,+0.40}$
& $\phantom{-}1.67_{\,-0.37}^{\,+0.42}$
& -- \\ \addlinespace
  $\bar{B}_s\to D_s^{+}K^{\ast-}$
& $\phantom{-}5.15$
& $\phantom{-}5.45_{\,-1.52}^{\,+1.74}$
& $\phantom{-}5.64_{\,-1.56}^{\,+1.79}$
& -- \\ \addlinespace
  $\bar{B}_s\to D_s^{\ast+}K^{\ast-}$
& $\phantom{-}3.02$
& $\phantom{-}3.19_{\,-0.65}^{\,+0.71}$
& $\phantom{-}3.31_{\,-0.67}^{\,+0.72}$
& -- \\ \addlinespace
\hline
&&&&\\[-0.3cm]
  $\bar{B}_d\to D^+a_1^-$
& $\phantom{0}10.82$
& $\phantom{0}11.44_{\,-1.48}^{\,+1.55}$
& $\phantom{0}11.84_{\,-1.50}^{\,+1.55}$
& $\phantom{-}6.0\pm3.3$ \\ \addlinespace
  $\bar{B}_d\to D^{\ast+}a_1^-$
& $\phantom{0}10.12$
& $\phantom{0}10.66_{\,-1.06}^{\,+1.11}$
& $\phantom{0}11.06_{\,-1.07}^{\,+1.10}$
& -- \\ \addlinespace
  $\bar{B}_s\to D_s^{+}a_1^{-}$
& $\phantom{0}11.23$
& $\phantom{0}11.87_{\,-3.36}^{\,+3.84}$
& $\phantom{0}12.29_{\,-3.46}^{\,+3.95}$
& -- \\ \addlinespace
  $\bar{B}_s\to D_s^{\ast+}a_1^{-}$
& $\phantom{-}7.44$
& $\phantom{-}7.84_{\,-1.53}^{\,+1.64}$
& $\phantom{-}8.13_{\,-1.57}^{\,+1.68}$
& -- \\ \addlinespace
\hline \hline
\end{tabular}
\end{center}
\end{table}

From Table~\ref{tab:br}, one can see that our predictions for the branching ratios of these decays generally come out higher than the experimental data, especially for $\bar{B}_{d}\to D^{(\ast)+}\pi^-$ and $\bar{B}_{d}\to D^{(\ast)+}\rho^-$ decays, where the difference in central values is at the 20~--~30\% level. Taking into account the uncertainties, the deviation is at the level of 2~--~3$\sig$. Compared to ref.~\cite{Beneke:2000ry}, which found at NLO rather good agreement between theory and experiment, essentially three things have changed: First, using the latest extraction from~\cite{Amhis:2014hma,Sakaki:2012ft,Fajfer:2012vx} our numerical values for the form factors are about 10\% larger than the ones used in~\cite{Beneke:2000ry}. Second, the NNLO corrections add another positive shift of 2~--~3\% on the amplitude level. Third, the experimental central values have slightly decreased since the analysis of ~\cite{Beneke:2000ry}. All three effects shift theory and experiment further apart.

Given the fact that the results show rough agreement within errors for $\bar{B}_{d}\to D^{(\ast)+}K^{(\ast)-}$ decays, which receive only contributions from colour-allowed tree topologies, this may indicate a non-negligible impact from the $W$-exchange topologies appearing only in $\bar{B}_{d}\to D^{(\ast)+}\pi^-$ and $\bar{B}_{d}\to D^{(\ast)+}\rho^-$ decays. For $\bar{B}_s$ decays, on the other hand, since the $B_s\to D_s^{(\ast)}$ transition form factors have so far received only little theoretical attention~\cite{Blasi:1993fi,Jenkins:1992qv,Li:2009wq,Li:2010bb,Chen:2011ut,Faustov:2012mt,Fan:2013kqa}, especially by the lattice QCD community~\cite{Bailey:2012rr,Atoui:2013zza}, our theoretical predictions are still plagued by larger uncertainties due to these hadronic parameters.

\subsection{Test of factorization}
\label{subsec:fact_test}

To further test the factorization hypothesis in class-I decays of $B$-mesons into heavy-light final states, as well as to probe the non-factorizable corrections to the coefficients $a_1(D^{(\ast)+}L^-)$, we now consider either ratios of non-leptonic to semi-leptonic decay rates~\cite{Beneke:2000ry,Bjorken:1988kk,Neubert:1997uc,Fleischer:2010ca}, or ratios of two non-leptonic decay rates~\cite{Beneke:2000ry,Neubert:1997uc}, both of which are essentially free of CKM and hadronic uncertainties.

As suggested firstly by Bjorken~\cite{Bjorken:1988kk}, a particularly clean and direct method to test the factorization hypothesis is provided by dividing the non-leptonic $\bar{B}_{d}\to D^{(\ast)+}L^-$ decay rates by the corresponding differential semi-leptonic $\bar{B}_{d}\to D^{(\ast)+}\ell^-\bar{\nu}_{\ell}$ decay rates evaluated at $q^2=m_L^2$, where $\ell$ refers to either an electron or a muon, and $q^2$ is the four-momentum squared transferring to the lepton pair. In this way, the coefficients $a_1(D^{(\ast)+}L^-)$ can be extracted directly from experimental data through the relation~\cite{Beneke:2000ry,Neubert:1997uc}
\begin{align} \label{eq:nonlep2semilep}
R_L^{(\ast)} &\equiv \frac{\Gamma(\bar{B}_{d}\to D^{(\ast)+}L^-)}{d\Gamma(\bar{B}_{d}\to D^{(\ast)+}\ell^-\bar{\nu}_{\ell})/dq^2\mid_{q^2=m_L^2}}\, = \, 6\pi^2\,|V_{ij}|^2\,f_L^2\,|a_1(D^{(\ast)+}L^-)|^2\, X_L^{(\ast)}\,,
\end{align}
where $V_{ij}$ is, depending on the constituent quark content of the meson $L$, the appropriate CKM matrix element. With the light lepton mass neglected, $X_L=X_L^{\ast}=1$ for a vector or axial-vector meson, whereas for a pseudoscalar $X_L^{(\ast)}$ deviates from unity only by calculable terms of order $m_L^2/m_B^2$, which are numerically below the percent level; explicit expressions for $X_L^{(\ast)}$ can be found, for example, in ref.~\cite{Neubert:1997uc}. To get the differential semi-leptonic decay rates at $q^2=m_L^2$ in eq.~\eqref{eq:nonlep2semilep}, we use the CLN parameterization for the $B\to D^{(\ast)}$ transition form factors~\cite{Caprini:1997mu}, with the relevant parameters summarized in Table~\ref{tab:inputs}. Explicitly, we get numerically~(in units of $10^{-3}~\mathrm{GeV}^{-2}~\mathrm{ps}^{-1}$)
\begin{align}
\frac{d\Gamma(\bar{B}_{d}\to D^{(\ast)+}\ell^-\bar{\nu}_{\ell})}{dq^2}\bigg\rvert_{q^2=m_L^2}
&=\left\{
      \begin{array}{ll}
       2.35^{+0.25}_{-0.24}~(2.04\pm0.10),  & \text{for $L=\pi^-$}\\[0.2cm]
       2.27^{+0.23}_{-0.22}~(2.28\pm0.10),  & \text{for $L=\rho^-$}\\[0.2cm]
       2.32^{+0.24}_{-0.23}~(2.14\pm0.10),  & \text{for $L=K^-$}\\[0.2cm]
       2.24^{+0.23}_{-0.22}~(2.36\pm0.10),  & \text{for $L=K^{\ast-}$}\\[0.2cm]
       2.13^{+0.21}_{-0.20}~(2.64\pm0.11),  & \text{for $L=a_1^-$}
      \end{array}
  \right. .
\end{align}
Together with the data on the branching ratios of non-leptonic decays given in Table~\ref{tab:br}, we arrive at the experimental values for $|a_1(D^{(\ast)+}L^-)|$ collected in Table~\ref{tab:nonlep2semilep}, where, for comparison, our theoretical predictions at different orders are also shown.

\begin{table}[t]
\tabcolsep0.53cm
\let\oldarraystretch=\arraystretch
\renewcommand*{\arraystretch}{1.1}
\begin{center}
\caption{\label{tab:nonlep2semilep} Theoretical predictions for $|a_1(D^{(\ast)+}L^-)|$ at different orders in perturbation theory. For comparison, the coefficients $|a_1(D^{(\ast)+}L^-)|$ determined from current data are shown in the last column. The experimental errors are estimated by adding the uncertainties of the non-leptonic branching ratios and the semi-leptonic decay rates in quadrature, while the uncertainties from the decay constants are not taken into account.}
\vspace*{0.16cm}
\begin{tabular}{lccccc}
\hline \hline
&&&&\\[-0.5cm]
$|a_1(D^{(\ast)+}L^-)|$ & ${\rm LO}$ & ${\rm NLO}$ & ${\rm NNLO}$ & Exp. \\
\hline
&&&&\\[-0.3cm]
  $|a_1(D^{+}\pi^-)|$
& $\phantom{-}1.025$
& $\phantom{-}1.054_{\,-0.020}^{\,+0.022}$
& $\phantom{-}1.073_{\,-0.014}^{\,+0.012}$
& $\phantom{-}0.89\pm0.05$ \\ \addlinespace
  $|a_1(D^{\ast+}\pi^-)|$
& $\phantom{-}1.025$
& $\phantom{-}1.052_{\,-0.018}^{\,+0.020}$
& $\phantom{-}1.071_{\,-0.014}^{\,+0.013}$
& $\phantom{-}0.96\pm0.03$ \\ \addlinespace
  $|a_1(D^{+}\rho^-)|$
& $\phantom{-}1.025$
& $\phantom{-}1.054_{\,-0.019}^{\,+0.022}$
& $\phantom{-}1.072_{\,-0.014}^{\,+0.012}$
& $\phantom{-}0.91\pm0.08$ \\ \addlinespace
  $|a_1(D^{\ast+}\rho^-)|$
& $\phantom{-}1.025$
& $\phantom{-}1.052_{\,-0.018}^{\,+0.020}$
& $\phantom{-}1.071_{\,-0.014}^{\,+0.013}$
& $\phantom{-}0.86\pm0.06$ \\ \addlinespace
\hline
&&&&\\[-0.3cm]
  $|a_1(D^{+}K^-)|$
& $\phantom{-}1.025$
& $\phantom{-}1.054_{\,-0.019}^{\,+0.022}$
& $\phantom{-}1.070_{\,-0.013}^{\,+0.010}$
& $\phantom{-}0.87\pm0.06$ \\ \addlinespace
  $|a_1(D^{\ast+}K^-)|$
& $\phantom{-}1.025$
& $\phantom{-}1.052_{\,-0.018}^{\,+0.020}$
& $\phantom{-}1.069_{\,-0.013}^{\,+0.010}$
& $\phantom{-}0.97\pm0.04$ \\ \addlinespace
  $|a_1(D^{+}K^{\ast-})|$
& $\phantom{-}1.025$
& $\phantom{-}1.054_{\,-0.019}^{\,+0.022}$
& $\phantom{-}1.070_{\,-0.013}^{\,+0.010}$
& $\phantom{-}0.99\pm0.09$ \\ \addlinespace
\hline
&&&&\\[-0.3cm]
  $|a_1(D^{+}a_1^-)|$
& $\phantom{-}1.025$
& $\phantom{-}1.054_{\,-0.019}^{\,+0.022}$
& $\phantom{-}1.072_{\,-0.014}^{\,+0.012}$
& $\phantom{-}0.76\pm0.19$ \\ \addlinespace
\hline \hline

\vspace*{-32pt}

\end{tabular}
\end{center}
\end{table}

From Table~\ref{tab:nonlep2semilep}, one can see clearly that our theoretical predictions based on the QCDF approach result in an essentially universal value of $|a_1(D^{(\ast)+}L^-)|\simeq 1.07~(1.05)$ at NNLO~(NLO), being consistently higher than the central values favoured by the current experimental data. The deviation is again at the level of 2~--~3$\sig$. Similar results were obtained in~\cite{Fleischer:2010ca}, yet without inclusion of the NNLO correction. It would be, therefore, very encouraging to determine directly the ratios of non-leptonic and semi-leptonic decay rates at current and future experiments such as LHCb and Belle~II. Compared to the NLO analysis in~\cite{Beneke:2000ry}, where theory predictions for $|a_1(D^{(\ast)+}L^-)|$  were found to be in agreement with the values extracted from experiment, together with the conclusion that there was no hint for sizable power corrections, the situation has changed, owing to increased values in the theory predictions and, at the same time, decreased experimental values~(see also the discussion in section~\ref{subsec:brs}). We will come back to this point below.

We now turn to discuss the ratios of non-leptonic $\bar{B}_{(s)}\to D_{(s)}^{(\ast)+}L^-$ decay rates, following the notations used in refs.~\cite{Beneke:2000ry,Neubert:1997uc}. As a quasi-universal $|a_1(D^{(\ast)+}L^-)|$ is predicted in the QCDF approach, these ratios could be used to test the factorization hypothesis, as well as the $\mathrm{SU(3)}$ relations in $B$-meson decays into heavy-light final states~\cite{Fleischer:2010ca}. Our results of such an analysis are presented in Table~\ref{tab:nonlepratio}, where the experimental data is obtained using the corresponding branching fractions collected in Table~\ref{tab:br}.

\begin{table}[t]
\tabcolsep0.48cm
\let\oldarraystretch=\arraystretch
\renewcommand*{\arraystretch}{1.1}
\begin{center}
\caption{\label{tab:nonlepratio} Predictions for the ratios of non-leptonic $\bar{B}_{(s)}\to D_{(s)}^{(\ast)+}L^-$ decay rates at different orders. The experimental data is obtained using the corresponding branching fractions collected in Table~\ref{tab:br}.}
\vspace{0.2cm}
\begin{tabular}{lccccc}
\hline \hline
&&&&\\[-0.5cm]
Ratios & ${\rm LO}$ & ${\rm NLO}$ & ${\rm NNLO}$ & Exp. \\
\hline
&&&&\\[-0.3cm]
  $\frac{{\rm Br}(\bar{B}_{d}\to D^{\ast+}\pi^-)}{
         {\rm Br}(\bar{B}_{d}\to D^{+}\pi^-)}$
& $\phantom{-}0.880$
& $\phantom{-}0.876_{\,-0.150}^{\,+0.162}$
& $\phantom{-}0.878_{\,-0.150}^{\,+0.162}$
& $\phantom{-}1.03\pm0.07$ \\ \addlinespace
  $\frac{{\rm Br}(\bar{B}_{d}\to D^{+}\rho^-)}{
         {\rm Br}(\bar{B}_{d}\to D^{+}\pi^-)}$
& $\phantom{-}2.654$
& $\phantom{-}2.653_{\,-0.158}^{\,+0.163}$
& $\phantom{-}2.653_{\,-0.158}^{\,+0.163}$
& $\phantom{-}2.80\pm0.47$ \\ \addlinespace
  $\frac{{\rm Br}(\bar{B}_{d}\to D^{+}\rho^-)}{
         {\rm Br}(\bar{B}_{d}\to D^{\ast+}\pi^-)}$
& $\phantom{-}3.016$
& $\phantom{-}3.027_{\,-0.531}^{\,+0.599}$
& $\phantom{-}3.022_{\,-0.530}^{\,+0.598}$
& $\phantom{-}2.72\pm0.45$ \\ \addlinespace
\hline
&&&&\\[-0.3cm]
  $\frac{{\rm Br}(\bar{B}_{d}\to D^{\ast+}K^-)}{
         {\rm Br}(\bar{B}_{d}\to D^{+}K^-)}$
& $\phantom{-}0.865$
& $\phantom{-}0.862_{\,-0.147}^{\,+0.158}$
& $\phantom{-}0.863_{\,-0.147}^{\,+0.158}$
& $\phantom{-}1.086\pm0.141$ \\ \addlinespace
  $\frac{{\rm Br}(\bar{B}_{d}\to D^{+}K^{\ast-})}{
         {\rm Br}(\bar{B}_{d}\to D^{+}K^-)}$
& $\phantom{-}1.747$
& $\phantom{-}1.746_{\,-0.115}^{\,+0.118}$
& $\phantom{-}1.746_{\,-0.115}^{\,+0.118}$
& $\phantom{-}2.284\pm0.431$ \\ \addlinespace
  $\frac{{\rm Br}(\bar{B}_{d}\to D^{+}K^{\ast-})}{
         {\rm Br}(\bar{B}_{d}\to D^{\ast+}K^-)}$
& $\phantom{-}2.019$
& $\phantom{-}2.026_{\,-0.358}^{\,+0.404}$
& $\phantom{-}2.023_{\,-0.358}^{\,+0.403}$
& $\phantom{-}2.103\pm0.363$ \\ \addlinespace
\hline
&&&&\\[-0.3cm]
  $\frac{{\rm Br}(\bar{B}_{d}\to D^{+}K^-)}{
         {\rm Br}(\bar{B}_{d}\to D^{+}\pi^-)}$
& $\phantom{-}0.077$
& $\phantom{-}0.077_{\,-0.002}^{\,+0.002}$
& $\phantom{-}0.077_{\,-0.002}^{\,+0.002}$
& $\phantom{-}0.074\pm0.009$ \\ \addlinespace
  $\frac{{\rm Br}(\bar{B}_{d}\to D^{\ast+}K^-)}{
         {\rm Br}(\bar{B}_{d}\to D^{\ast+}\pi^-)}$
& $\phantom{-}0.075$
& $\phantom{-}0.075_{\,-0.002}^{\,+0.002}$
& $\phantom{-}0.075_{\,-0.002}^{\,+0.002}$
& $\phantom{-}0.078\pm0.007$ \\ \addlinespace
  $\frac{{\rm Br}(\bar{B}_{d}\to D^{+}K^{\ast-})}{
         {\rm Br}(\bar{B}_{d}\to D^{+}\rho^-)}$
& $\phantom{-}0.050$
& $\phantom{-}0.050_{\,-0.004}^{\,+0.005}$
& $\phantom{-}0.050_{\,-0.004}^{\,+0.005}$
& $\phantom{-}0.060\pm0.013$ \\ \addlinespace
\hline
&&&&\\[-0.3cm]
  $\frac{{\rm Br}(\bar{B}_{s}\to D_{s}^{+}\pi^-)}{
         {\rm Br}(\bar{B}_{d}\to D^{+}K^-)}$
& $\phantom{-}14.67$
& $\phantom{-}14.67_{\,-1.28}^{\,+1.34}$
& $\phantom{-}14.67_{\,-1.28}^{\,+1.34}$
& $\phantom{-}15.43\pm2.02$ \\ \addlinespace
  $\frac{{\rm Br}(\bar{B}_{s}\to D_{s}^{+}\pi^-)}{
         {\rm Br}(\bar{B}_{d}\to D^{+}\pi^-)}$
& $\phantom{-}1.120$
& $\phantom{-}1.120_{\,-0.104}^{\,+0.109}$
& $\phantom{-}1.120_{\,-0.104}^{\,+0.109}$
& $\phantom{-}1.134\pm0.102$ \\ \addlinespace
\hline \hline

\vspace*{-32pt}

\end{tabular}
\end{center}
\end{table}

From Table~\ref{tab:nonlepratio}, one can see that, within the errors, our theoretical predictions are generally consistent with the current experimental data, indicating therefore no evidence for any significant deviation from the factorization hypothesis for these class-I $B$-meson decays into heavy-light final states. The last two ratios shown in Table~\ref{tab:nonlepratio} could also be used to determine the ratio of fragmentation functions $f_d/f_s$, a key quantity for precise measurements of absolute $B_s$-meson decay rates at hadron colliders~\cite{Fleischer:2010ca,Fleischer:2010ay}.

One possible interpretation of our findings that, on the one hand, the non-leptonic to semi-leptonic ratios come out larger in theory compared to experiment, and on the other hand the non-leptonic ratios in general agree with experiment, might be non-negligible power corrections which could be negative in sign and 10~--~15\% in size on the amplitude level. They would render the factorization test via non-leptonic to semi-leptonic ratios better, and at the same time could cancel out in the non-leptonic ratios, especially if they were of a certain universality. The size of power corrections stemming from spectator scattering and weak annihilation was roughly estimated in section 6.5 of~\cite{Beneke:2000ry}. Depending on the phases of the integrals over the $D$-meson wave function and on the value of the first inverse moment $\lambda_B$ of the $B$-meson distribution amplitude, these two contributions could in principle interfere constructively, and in this case their total effect could indeed add up to $-10$\% in the amplitude.

Another possibility which has essentially the same effect would be to reduce the values of $|V_{cb}|$ times the form factors by $\sim 10$\%. This option seems attractive in view of the fact that those non-leptonic ratios in Table~\ref{tab:nonlepratio} in which $|V_{cb}|$ and the form factors cancel out are in very good agreement with experiment. On the other hand, the semi-leptonic rate is measured very precisely and the current form factors times $|V_{cb}|$ are extracted by HFAG~\cite{Amhis:2014hma} from a global fit to all available data, whose result we quote in Table~\ref{tab:inputs}. Hence they are optimized to describe the shape of the semi-leptonic rate and therefore should be trustworthy. One could even conclude from this that the experimental extraction of $|a_1(D^{(\ast)+}L^-)|$ from eq.~\eqref{eq:nonlep2semilep} is independent of the product of $|V_{cb}|$ and the form factor.

We emphasize that without a rigorous treatment of power corrections in the QCDF approach nothing more can be said at the present stage. In any case, the QCDF approach per se is not invalidated.

\subsection{Predictions for $\Lambda_b \to \Lambda_c^+ L^-$ decays}
\label{subsec:lambdab}

While the $\Lambda_b$ baryons are not produced at an $e^+ e^-$ $B$-factory, they account for about $20\%$ of the $b$-hadrons produced at the LHC~\cite{Aaij:2011jp}. Remarkably, the number of $\Lambda_b$ baryons produced is comparable to the number of $B_u$ or $B_d$ mesons, and is significantly higher than the number of $B_s$ mesons. Due to the half-integer-spin of $\Lambda_b$, its decays provide complementary information compared to the corresponding mesonic ones. Therefore, this may open up a new field for flavour physics. For a review, see e.g.\ refs.~\cite{Korner:1994nh,Klempt:2009pi,Meinel:2014wua,Wang:2014jya}. Here we study the two-body non-leptonic $\Lambda_b \to \Lambda_c^+ L^-$ decays, for which the factorization assumption is believed to be reliable~\cite{Korner:1992wi,Cheng:1996cs,Ivanov:1997ra,Leibovich:2003tw,Ke:2007tg}. As demonstrated especially in ref.~\cite{Leibovich:2003tw}, the proof of factorization at leading order in $\Lambda_{\rm QCD}/m_{b,c}$ for these decays follows closely that for $\bar{B}_d \to D^{
(\ast)+}\pi^{-}$~\cite{Bauer:2001cu}. These decays provide, therefore, a testing ground for different QCD models and factorization assumptions used in $B$-meson case. It is straightforward to generalize the expressions in~\cite{Leibovich:2003tw} to take radiative corrections through to NNLO into account.

\begin{table}[t]
\tabcolsep0.50cm
\let\oldarraystretch=\arraystretch
\renewcommand*{\arraystretch}{1.1}
\begin{center}
\caption{\label{tab:baryon} Predictions for the branching fractions (in units of $10^{-3}$ for $b\to c\bar{u}d$ and $10^{-4}$ for $b\to c\bar{u}s$ transitions) of $\Lambda_b \to \Lambda_c^+ L^-$ decays, as well as some ratios between them. The experimental data is taken from PDG~\cite{Agashe:2014kda} and HFAG~\cite{Amhis:2014hma}.}
\vspace{0.2cm}
\begin{tabular}{lccccc}
\hline \hline
&&&&\\[-0.5cm]
Decay mode & ${\rm LO}$ & ${\rm NLO}$ & ${\rm NNLO}$ & Exp. \\
\hline
&&&&\\[-0.3cm]
  $\Lambda_b \to \Lambda_c^+ \pi^-$
& $\phantom{-}2.60$
& $\phantom{-}2.75_{\,-0.53}^{\,+0.53}$
& $\phantom{-}2.85_{\,-0.54}^{\,+0.54}$
& $\phantom{-}4.30_{\,-0.35}^{\,+0.36}$ \\ \addlinespace
  $\Lambda_b \to \Lambda_c^+ \rho^-$
& $\phantom{-}7.46$
& $\phantom{-}7.88_{\,-1.43}^{\,+1.44}$
& $\phantom{-}8.17_{\,-1.47}^{\,+1.47}$
& -- \\ \addlinespace
  $\Lambda_b \to \Lambda_c^+ a_1^{-}$
& $\phantom{-}9.57$
& $\phantom{0}10.11_{\,-1.72}^{\,+1.75}$
& $\phantom{0}10.47_{\,-1.77}^{\,+1.78}$
& -- \\ \addlinespace
\hline
&&&&\\[-0.3cm]
  $\Lambda_b \to \Lambda_c^+ K^-$
& $\phantom{-}2.02$
& $\phantom{-}2.14_{\,-0.39}^{\,+0.40}$
& $\phantom{-}2.21_{\,-0.40}^{\,+0.40}$
& $\phantom{-}3.42\pm0.33$ \\ \addlinespace
  $\Lambda_b \to \Lambda_c^+ K^{\ast-}$
& $\phantom{-}3.86$
& $\phantom{-}4.07_{\,-0.73}^{\,+0.74}$
& $\phantom{-}4.22_{\,-0.75}^{\,+0.75}$
& -- \\ \addlinespace
\hline
&&&&\\[-0.3cm]
  $\frac{{\rm Br}(\Lambda_b\to \Lambda_c^{+} \mu^{-} \bar{\nu})}{{\rm Br}(\Lambda_b\to \Lambda_c^{+} \pi^{-})}$
& $\phantom{-}18.88$
& $\phantom{-}17.87_{\,-2.33}^{\,+2.31}$
& $\phantom{-}17.25_{\,-2.18}^{\,+2.19}$
& $\phantom{-}16.6_{\,-4.7}^{\,+4.1}$ \\ \addlinespace
  $\frac{{\rm Br}(\Lambda_b\to \Lambda_c^{+} K^{-})}{{\rm Br}(\Lambda_b\to \Lambda_c^{+} \pi^{-})}~(\%)$
& $\phantom{-}7.77$
& $\phantom{-}7.77_{\,-0.18}^{\,+0.19}$
& $\phantom{-}7.77_{\,-0.18}^{\,+0.19}$
& $\phantom{-}7.31\pm0.23$ \\ \addlinespace
  $\frac{{\rm Br}(\Lambda_b\to \Lambda_c^{+} \pi^{-})}{{\rm Br}(\bar{B}_{d}\to D^{+} \pi^{-})}$
& $\phantom{-}0.73$
& $\phantom{-}0.73_{\,-0.15}^{\,+0.16}$
& $\phantom{-}0.73_{\,-0.15}^{\,+0.16}$
& $\phantom{-}3.3\pm1.2$ \\ \addlinespace
\hline \hline

\vspace*{-32pt}

\end{tabular}
\end{center}
\end{table}

Using the most recent lattice QCD results for $\Lambda_b \to \Lambda_c$ transition form factors~\cite{Detmold:2015aaa}, we present in Table~\ref{tab:baryon} our predictions for the branching fractions of $\Lambda_b \to \Lambda_c^+ L^-$ decays, as well as some ratios between them, where the experimental data is taken from HFAG~\cite{Amhis:2014hma}. From Table~\ref{tab:baryon}, one can see that, contrary to the observation made in mesonic decays, our predictions for the branching ratios of these decays now come out lower than the experimental data; especially the higher-order corrections always increase the LO predictions and shift our predictions closer to the experimental data. Our predictions for the two ratios ${\rm Br}(\Lambda_b\to \Lambda_c^{+} \mu^{-} \bar{\nu})/{\rm Br}(\Lambda_b\to \Lambda_c^{+} \pi^{-})$ and ${\rm Br}(\Lambda_b\to \Lambda_c^{+} K^{-})/{\rm Br}(\Lambda_b\to \Lambda_c^{+} \pi^{-})$ are both consistent with the current data, indicating that the non-factorizable effects
should be small in these decays. Moreover, we emphasize the fact that the non-leptonic to semi-leptonic ratio in the baryonic case is consistent with experiment, but shows a tension in the mesonic case (see section~\ref{subsec:fact_test}). The discrepancy between our prediction and the current experimental data for ${\rm Br}(\Lambda_b\to \Lambda_c^{+} \pi^{-})/{\rm Br}(\bar{B}_{d}\to D^{+} \pi^{-})$ makes it interesting to evaluate directly the form-factor ratios of $\Lambda_b\to \Lambda_c$ and $B \to D$ transitions by the lattice community.

\section{Conclusion}
\label{sec:conclusion}

We have calculated the NNLO vertex corrections to the colour-allowed tree topology in the framework of QCDF for the mesonic decays $\bar{B}_{(s)}\to D_{(s)}^{(\ast)+} \, L^-$ and the baryonic decays $\Lambda_b \to \Lambda_c^+ \, L^-$, with $L=\{\pi,\rho,K^{(\ast)},a_1\}$. The calculation of the two-loop correction to the hard scattering kernels requires the evaluation of several dozens of genuine two-scale Feynman diagrams, which describe these heavy-to-heavy transitions at the quark level. We performed this calculation by means of techniques that have become standard in the business of multi-loop computations. It might be worth noting that we evaluated all master integrals analytically~\cite{Huber:2015bva} in a so-called canonical basis~\cite{Henn:2013pwa}, a result which catalyzed the convolution with the LCDA and enabled us to obtain the convoluted kernels almost completely analytically.

The NNLO contributions yield a positive shift to the colour-allowed tree amplitude $a_1$, which is sizable for its imaginary part, but small for its real part and its magnitude. Moreover, the amplitude only mildly depends on the ratio of the heavy-quark masses $z_c=m_c^2/m_b^2$. The dependence on the factorization scale gets reduced for the real part compared to the NLO result. This reduction does not occur in the imaginary parts, which is expected, as the latter only arise beyond LO. We performed our analysis using the pole scheme for the heavy-quark masses. A change to the $\overline{\text{MS}}$ scheme does not show any significant shift of the amplitude within the range of physical values for the heavy-quark masses. Moreover, the results for the different final states only slightly depend on the light meson LCDA and hence, we can confirm the quasi-universality of the tree amplitude to NNLO accuracy.

In our phenomenological analysis we evaluated the branching ratios to NNLO accuracy, and with the latest values for the non-perturbative input parameters. We find that for $\bar B_d$ decays the central values of the theoretical predictions are in general higher compared to the experimental values. Within the given uncertainties the quantities agree at the 2~--~3$\sig$ level for $\pi$ and $\rho$ in the final state, and slightly better for $K$ and $K^\ast$. Compared to the analysis at NLO~\cite{Beneke:2000ry}, our increased values for the form factors and the amplitude, together with decreased experimental values, have shifted theory and experiment further apart. For $\bar B_s$ decays, the theory predictions are still plagued by large uncertainties which are mainly due to poorly known form factors. For the baryonic decays, on the other hand, the predicted branching fractions turn out to be $20-30\%$ smaller than the experimental ones. It would be interesting to understand the reason for this difference in the $\bar B_d$ and the $\Lambda_b$ decays. We therefore propose a systematic analysis of factorization for $\Lambda_b$ decays in the future.

Moreover, we analyzed ratios of non-leptonic and semi-leptonic decay rates in order to further probe the factorization theorem to NNLO.
The ratios of different non-leptonic rates turn out to be in good agreement, comparing theoretical prediction and experiment.
In the case of non-leptonic to semi-leptonic ratios, on the other hand, the values for $|a_1(D^{(*)+}L^-)|$ that we extract from experiment are lower by 2~--~3$\sig$ compared to the NNLO theory predictions (see also~\cite{Fleischer:2010ca}).

One possibility to interpret the entity of these results could be non-negligible power corrections.
Given the uncertainties of the branching ratios and $a_1$ they could be negative in sign and 10~--~15\% in size on the amplitude level. They could cure the
non-leptonic to semi-leptonic ratios, without destroying the agreement in the non-leptonic ratios, especially if they were of a certain universality. It will also be very interesting to investigate what this would imply for the power corrections in charmless non-leptonic decays. Recent analyses that address weak annihilation in charmless non-leptonic decays can be found in~\cite{Bobeth:2014rra,Wang:2013fya,Chang:2014yma}.

Another, yet less favourable option would be reduced values of $|V_{cb}|$ times the form factors. As stated already in section~\ref{sec:pheno}, without a rigorous treatment of power corrections in the QCDF approach nothing more can be said at the present stage.

\section*{Acknowledgements}
We would like to thank Martin Beneke and Oleg Tarasov for collaboration at an initial stage of this project.
Moreover, we would like to thank Guido Bell, Martin Beneke, Thorsten Feldmann, and Bj\"orn~O.~Lange for very helpful discussions, and
Guido Bell and Martin Beneke for comments on the manuscript. This work was supported in part by the NNSFC of China under
contract Nos.~11675061 and 11435003~(XL), and by DFG Forschergruppe
FOR 1873 ``Quark Flavour Physics and Effective Field Theories''~(TH and SK). XL is also supported
in part by the SRF for ROCS, SEM, and by the self-determined research funds of CCNU from the
colleges' basic research and operation of MOE (CCNU15A02037).
XL acknowledges hospitality from Siegen University during the final stages of this work.


\end{document}